
\def\nt{\noindent}
\def\ce{\centerline}

\def\section#1#2{\vskip1truecm\nt{\bf#1\ #2}\vskip0.5truecm}\indent
\def\subsection#1#2{\vskip1truecm\nt{\bf#1\ #2}\vskip0.5truecm}\indent
\def\subsubsection#1#2{\vskip1truecm\nt{\bf#1\ #2}\vskip0.5truecm}\indent
\def\entry#1{\vskip1truecm\nt{\bf#1}\vskip0.5truecm}\indent

\magnification=1200
\normalbaselines
\nopagenumbers

\hsize=6.5 true in
\vsize=9.0 true in

\ce{\bf CONFORMAL FOURTH-RANK GRAVITY}

\bigskip

\ce{{\bf Victor Tapia}\footnote{*}{e-mail: VTAPIA@HALCON.DPI.UDEC.CL}}

\medskip

\ce{Departamento de F{\'\i}sica}

\ce{Facultad de Ciencias F{\'\i}sicas y Matem\'aticas}

\ce{Universidad de Concepci\'on}

\ce{Casilla 3-C}

\ce{Concepci\'on, Chile}

\bigskip

\ce{\bf Az-Eddine Marrakchi}

\medskip

\ce{Laboratoire de Physique Th\'eorique}

\ce{Facult\'e des Sciences}

\ce{Rabat, Morocco}

\medskip

\ce{and}

\medskip

\ce{D\'epartement de Physique\footnote{**}{Permanent address.}}

\ce{Facult\'e des Sciences}

\ce{B. P. 1796 - Atlas}

\ce{Fes, Morocco}

\bigskip

\ce{and}

\bigskip

\ce{{\bf Mauricio Cataldo}\footnote{***}{e-mail:
YCATALDO@UBIOBIO.DCI.UBIOBIO.CL}}

\medskip

{\ce{Departamento de F{\'\i}sica}}

{\ce{Universidad del Bio Bio}}

{\ce{Casilla 5-C}}

{\ce{Concepci\'on, Chile}}

\vfill\eject

\nt{\bf Abstract.} We consider the consequences of describing the metric
properties of space-time through a quartic line element
$ds^4=G_{\mu\nu\lambda\rho}dx^\mu dx^\nu dx^\lambda dx^\rho$. The associated
"metric" is a fourth-rank tensor $G_{\mu\nu\lambda\rho}$. We construct a
theory for the gravitational field based on the fourth-rank metric
$G_{\mu\nu\lambda\rho}$ which is conformally invariant in four dimensions.
In the absence of matter the fourth-rank metric becomes of the form
$G_{\mu\nu\lambda\rho}=g_{(\mu\nu}g_{\lambda\rho )}$ therefore we recover a
Riemannian behaviour of the geometry; furthermore, the theory coincides with
General Relativity. In the presence of matter we can keep Riemannianicity,
but now gravitation couples in a different way to matter as compared to
General Relativity. We develop a simple cosmological model based on a FRW
metric with matter described by a perfect fluid. Our field equations predict
that the entropy is an increasing function of time. For $k_{obs}=0$ the field
equations predict $\Omega\approx 4y$, where $y={p\over\rho}$; for
$\Omega_{small}=0.01$ we obtain $y_{pred}=2.5\times10^{-3}$. $y$ can be
estimated from the mean random velocity of typical galaxies to be
$y_{random}=1\times10^{-5}$. For the early universe there is no violation of
causality for $t>t_{class}\approx10^{19}t_{Planck}\approx10^{-24}s$.

\bigskip

\nt Short title: Conformal fourth-rank gravity.

\bigskip

\nt Classification Number: 0450 Unified field theories and other theories of
gravitation.

\vfill\eject

{\it "The next case in simplicity includes those manifoldnesses in which the
line-element may be expressed as the fourth-root of a quartic differential
expression."}

\medskip

\rightline{\it B. Riemann, 1854}

\bigskip

\section{1.}{Introduction}

If we adopt the materialist vision that the physical world is an objective
reality then, necessarily, our geometrical conception of the universe is
limited by our psychological perception of it. There is in fact a
self-consistency in that physical laws generate the very mathematics
necessary to make those laws understandable. In other words, we can conceive
what nature allows us to conceive. In the scale of distances of our daily
life, i.e., distances much greater than the Planck length, the universe
behaves quite smoothly and one hopes that this behaviour might be
extrapolated to very large, cosmological, and also to very small, even
subnuclear, distances. This smooth behaviour would allow the universe to be
mathematically modeled by a differentiable manifold. Of course, the very
concept of a differentiable manifold is possible only because our perception
of space allows us to conceive it, and one can wonder how our mathematical
conceptions are restricted by this kind of anthropic principle.

It seems that the problem of determining the geometry realised in nature was
first addressed by Riemann$^1$ in his famous, but little read, thesis in
1854. He pointed out that this geometry has to be determined by purely
empirical, experimental and observational, means and cannot be decided upon
{\it a priori}. The first indirect statements about the metrical properties
of our universe can be found in the Pythagoras theorem which, in a modern
language, is equivalent to Riemannian geometry

$${ds}^2\,=\,g_{\mu\nu}(x)\,dx^\mu\,dx^\nu\,.\eqno{(1.1)}$$

\nt The only thing we can try to understand now is the Riemannian, or
Pythagorean, nature of the geometry. Here we take recourse to the classical
argumentation by Riemann.$^1$ The infinitesimal element of distance $ds$
should be a function of the coordinates $x$'s and their differentials $dx$'s

$$ds\,=\,(x,\,dx)\,.\eqno{(1.2)}$$

\nt This function must satisfy the single requirement

$$f(x,\,\lambda\,dx)\,=\,\mid\lambda\mid\,\,f(x,\,dx)\,.\eqno{(1.3)}$$

\nt Of course, the possibilities are infinitely many. Let us restrict our
considerations to monomial functions

$$ds\,=\,{(G_{\mu_1\cdots\mu_r}\,dx^{\mu_1}\,\cdots\,dx^{\mu_r})}^{1/r}\,.
\eqno{(1.4)}$$

\nt In order for this quantity to satisfy (1.3) $r$ must be an even number.
The simplest choice is $r=2$, which corresponds to Riemannian geometry.

As pointed out by Riemann, the next possibility is $r=4$. In this case the
line element is given by

$${ds}^4\,=\,G_{\mu\nu\lambda\rho}\,dx^\mu\,dx^\nu\,dx^\lambda\,dx^\rho\,.
\eqno{(1.5)}$$

\nt Riemann went no farther in exploring the above geometry, and gave no
justification for that omission. Of course, at first sight, a space with a
line element of the form (1.5) may seem bizarre. However, such geometry
cannot be excluded {\it a priori} and its exclusion must be done in a
mathematically educated way.

This was partially done by Helmholtz.$^2$ He showed that the existence of
rigid bodies, which do not change their shapes and therefore the metric
relations under translations and rotations, leaves us with Riemannian
geometry as the only possibility. The Helmholtz result seemed quite
satisfactory and therefore no more concern for higher-rank geometries
appeared. It seems that the arrival of General Relativity, with its
underlying Riemannian geometry, caused this important problem to be
forgotten. However, the problem merits further attention, not only from a
mathematical point of view, but also for the applications it found in
theoretical and mathematical physics. It is here that the introductory
considerations come into play. In fact, the difficulty of conceiving
geometries other than the Riemannian limited their developments.
We are therefore going to develop that chapter of differential geometry in
which Riemann and Helmholtz stopped their scientific enquiries.

To close these historical comments. An indirect verification of the
Riemannian structure of the universe at our daily life scales was performed
by Gauss$^3$ in 1826. The experiment was intended to verify departures from
flatness, but as a side result he also verified no departures from
Riemannianicity.

At the scale of distances of our daily life fourth-rank geometry is not
realised in nature and the only place where it can play some role is in
high-energy, or short distances, physics. In fact, at high energies, a regime
to which we do not have direct experimental access, the very concept of rigid
body may be no longer valid and the Helmholtz argumentation no longer
applicable.

The natural question now is: why we would like to work with fourth-rank
geometry, and no other of the infinitely many possible generalisations of
Riemannian geometry, to describe the physics at high energies. The answer is
provided by experiments, such as deep inelastic scattering, which show that,
at very high energies, physical processes are scale, or conformally,
invariant. Therefore, high-energy physics is associated to a geometry
exhibiting, in a model independent way, conformal invariance in 4 dimensions.
In another work$^4$ we show that the critical dimension, for which field
theories are integrable, is equal to the rank of the metric. Therefore, if we
want to construct an integrable field theory in 4 dimensions showing
agreement with the observed conformal invariance at high energies we must
take recourse to fourth-rank geometry. This result also explains why, if one
relies only on Riemannian geometry, integrable conformal models can be
constructed only in 2 dimensions (strings).

We arrive therefore to the following scheme: at short distances,
high-energies, the geometry is of fourth-rank while at large distances, low
energies, the geometry is of second-rank, Riemannian. It is clear furthermore
that the Riemannian behaviour of  the  geometry  must  be  recovered  as  the
low-energy limit of the high-energy theory. This  would  be  possible  if  at
low-energies the fourth-rank metric tensor $G_{\mu\nu\lambda\rho}$ becomes of
the form

$$G_{\mu\nu\lambda\rho}\,=\,g_{(\mu\nu}\,g_{\lambda\rho )}\,.\eqno{(1.6a)}$$

\nt In this case the line element factors and one is back to the Riemannian
case

$${ds}^4\,=\,{({ds}^2)}^2\,.\eqno{(1.6b)}$$

Our next task is to construct a geometric invariant to be used as the
Lagrangian describing the dynamics of the geometry, i.e., of the
gravitational field. From the metric alone it is imposible to construct any
invariant, apart from the trivial solution: a constant. Therefore, we must
take recourse to a further geometrical object: the Ricci tensor for an
arbitrary connection ${\Gamma^\lambda}_{\mu\nu}$

$$R_{\mu\nu}\,=\,\partial_\lambda{\Gamma^\lambda}_{\mu\nu}\,-\,\partial_\nu
{\Gamma^\lambda}_{\lambda\mu}\,+\,{\Gamma^\lambda}_{\lambda\sigma}\,{\Gamma^
\sigma}_{\mu\nu}\,-\,{\Gamma^\lambda}_{\mu\sigma}\,{\Gamma^\sigma}_{\lambda
\nu}\,.\eqno{(1.7)}$$

\nt The simplest invariants which can be constructed with the metric
$G_{\mu\nu\lambda\rho}$ and the Ricci tensor $R_{\mu\nu}$ are

$$\langle R^2\rangle\,=\,G^{\mu\nu\lambda\rho}\,R_{\mu\nu}\,R_{\lambda\rho}
\,,$$

$$\langle R^4\rangle\,=\,G^{\mu\nu\lambda\rho}\,G^{\alpha\beta\gamma\delta}\,
R_{\mu\alpha}\,R_{\nu\beta}\,R_{\lambda\gamma}\,R_{\rho\delta}\,,\quad etc.
\eqno{(1.8)}$$

\nt The Lagrangian therefore will be of the form

$${\cal L}\,=\,L(\langle R^2\rangle,\,\langle R^4\rangle,\,\cdots)\,G^{1/4}
\,,\eqno{(1.9)}$$

\nt where $G$ is the determinant of $G_{\mu\nu\lambda\rho}$. The scalar
function $L$ to be put in (1.9) should make the Lagrangian a conformally
invariant function. Under rescalings of the metric

$$G_{\mu\nu\lambda\rho}\,\rightarrow\,\lambda\,G_{\mu\nu\lambda\rho}\,,
\eqno{(1.10)}$$

\nt the inverse metric $G^{\mu\nu\lambda\rho}$ and $G^{1/4}$ transform as

$$G^{\mu\nu\lambda\rho}\,\rightarrow\,{\lambda}^{-1}\,G^{\mu\nu\lambda\rho}
\,,\eqno{(1.11a)}$$

$$G^{1/4}\,\rightarrow\,\lambda\,G^{1/4}\,.\eqno{(1.11b)}$$

\nt Therefore the Lagrangian should be of the form

$${\cal L}\,=\,[\alpha\,\langle R^2\rangle\,+\,\beta\,{{\langle R^4\rangle}
\over{\langle R^2\rangle}}+\cdots]\,G^{1/4}\,.\eqno{(1.12)}$$

\nt However, all the terms after the first one, are highly non-local.
Therefore, the only sensible solution is

$${\cal L}\,=\,\kappa_{CG}\,\langle R^2\rangle\,G^{1/4}\,,\eqno{(1.13)}$$

\nt where

$$\kappa_{CG}\,\approx\,\kappa_E\,{L_{Planck}}^2\,=\,{{\hbar c}\over{8\pi}}\,
,\eqno{(1.14)}$$

\nt is the Einstein gravitational constant $\kappa_E={c^4\over{8\pi G}}$,
times a constant of the order of ${L_{Planck}}^2$.

The total Lagrangian must also consider the contributions of matter. Now we
must apply a Palatini-like variational principle in which the connection and
the metric are varied independently. However, in all known cases of physical
interest the matter Lagrangian does not depend on the affine connection.$^5$
In this case the variation of the gravitational Lagrangian with respect to
the connection leads to a metricity condition for which the solution is

$${\Gamma^\lambda}_{\mu\nu}\,=\,\lbrace^\lambda_{\mu\nu}\rbrace(\gamma)\,.
\eqno{(1.15)}$$

\nt I.e., the connection is the Christoffel symbol of the second kind of the
tensor $\gamma^{\mu\nu}$ given by

$$\gamma^{\mu\nu}=G^{\mu\nu\lambda\rho}\,R_{\lambda\rho}\,,\eqno{(1.16)}$$

\nt which we have assumed to be regular. Equations (1.15) and (1.16) are a
metricity condition since they give the relation between
${\Gamma^\lambda}_{\mu\nu}$ and $G_{\mu\nu\lambda\rho}$. Therefore

$$R_{\mu\nu}(\Gamma)\,=\,R_{\mu\nu}(\gamma)\,.\eqno{(1.17)}$$

\nt One easily verifies then that

$$\langle R^2\rangle\,=\,G^{\mu\nu\lambda\rho}\,R_{\mu\nu}\,R_{\lambda\rho}\,
=\,\gamma^{\mu\nu}\,R_{\mu\nu}(\gamma)\,=\,R(\gamma)\,.\eqno{(1.18)}$$

Variation of the Lagrangian with respect to the metric
$G_{\mu\nu\lambda\rho}$ gives

$$\kappa_{CG}\,[ R_{(\mu\nu}\,R_{\lambda\rho)}\,-\,{1\over4}\,\langle R^2
\rangle\,G_{\mu\nu\lambda\rho}]\,=\,T_{\mu\nu\lambda\rho}\,,\eqno{(1.19)}$$

\nt where $T_{\mu\nu\lambda\rho}$ is the energy-momentum tensor of matter,
to be defined below.

The field equations (1.19) exhibit three energy regimes: low, medium, and
high. In the low-energy regime there is no matter and therefore the
fourth-rank metric is separable, $G_{\mu\nu\lambda\rho}=g_{(\mu\nu}
g_{\lambda\rho)}$, as can be read from (1.19). Then the line element would
factor, as in (1.6b), and one would be back to the Riemannian case. In the
medium-energy regime the geometry is still Riemannian, $G_{\mu\nu\lambda\rho
}=g_{(\mu\nu}g_{\lambda\rho)}$, but there is matter  involved  in  the  game.
This possibility is not excluded as a closer analysis of eqs. (1.19) reveals.
In this case the gravitational field couples in a different way, as  compared
to General Relativity, to matter. Lastly, we have the true high-energy regime
in which there is matter and the geometry is truly fourth-rank.

Let us further analyse these energy regimes. In vacuum, the field equations
(1.19) are equivalent to

$$R_{\mu\nu}(\gamma)\,-\,{1\over4}\,R(\gamma)\,\gamma_{\mu\nu}\,=0\,.
\eqno{(1.20)}$$

\nt For a spherically symmetric field the solution is the Kottler metric$^6$
which contains the Schwarzschild solution as a special case. Therefore the
predictions based on the \break Schwarzschild metric, which agree with
observation by 1 per cent or better, will be contained in this theory.

The large scale geometry of the universe seems to be Riemannian and, since
there is matter present in it, this corresponds to the medium-energy regime
mentioned above. In this context we develop a cosmological model based on the
Friedman-Robertson-Walker metric coupled to cosmic matter described by a
perfect fluid.

The theory predicts an increasing total entropy such that the expansion of
the universe is an adiabatic non-isoentropic process. Therefore, the
evolution of the universe, in the framework of fourth-rank cosmology is, as
expected on physical grounds, an irreversible process.

For $k_{obs}=0$, as imposed by the observed flatness of the universe, the
field equations give, for the present Universe,

$$\Omega\,=\,{4y\over{1-4y-y^2}}\,,\eqno{(1.21)}$$

\nt where $y={p\over\rho}$. For $\Omega_{small}=0.01$ [7] we obtain $y_{pred}
\approx2.5\times10^{-3}$ which corresponds to an almost pressureless  perfect
fluid. This must be compared with the observed value of  $p\over\rho$,  which
can be determined from the mean random velocity of typical  galaxies  and  is
given by $y_{random}=1\times10^{-5}$. Therefore, our  prediction  differs  by
two orders of magnitude with respect  to  the  observed  value.  We  hope  to
improve this situation since the estimation of $y$ from the random motion  of
galaxies is a quite rough one. Furthermore, eq. (1.21) was obtained under the
strong asumption that $y$ behaves like a constant. Therefore, there are hopes
that this theory shows a better agreement with the  observed  values  of  the
cosmological parameters.

For the early universe we find that causality is not violated for $t>t_{class
}\approx10^{19}t_{Planck}$ $\approx10^{-24}s$. At earlier times quantum
mechanical effects dominate the scene. In fact, the radius of the universe is
exactly the Compton wavelength associated to its mass. Our classical approach
breaks down so that the very concept of causality is meaningless.  Therefore,
there is no violation of causality, or horizon problem.

Some final introductory comments. It is a popular view that the gravitational
field is correctly described by General Relativity. This is true of the pure
gravitational field, i.e., when no coupling to matter, or other fields, is
present. In fact, Einstein field equations are in excellent agreement, 1 per
cent or better, with observation when applied, for example, to the solar
system. However, when matter is coupled to gravity the observational
agreement is not so good. This is the case when General Relativity is applied
to cosmology where the gravitational field get coupled to cosmic matter
described by a perfect fluid. One obtains qualitatively good predictions, as
the evolution of the universe from an initial singularity and some good
quantitative predictions as the temperature of the microwave background and
the relative abundance of elements. However, the quantitative agreement is
weaker in other aspects. In fact, flatness, $k_{obs}=0$, implies
$\Omega_{GR}=1$, which is hardly observed. Furthermore, the Standard Model of
Cosmology predicts a constant entropy, something which is difficult to accept
on physical grounds. These are some of the reasons to look for an improved
theory for the gravitational field.

In previous works$^{10,11,12}$ we developed a similar model based on the
Lagrangian

$${\cal L}\,=\,\kappa_E\,{\langle R^2\rangle}^{1\slash2}\,G^{1\slash4}\,.
\eqno{(1.22)}$$

\nt This Lagrangian was chosen in order to have only the Einstein
gravitational constant for dimensional purposes. Later on we became convinced
that the appearance of h in the Lagrangian (1.13) creates no conflict between
the classical character of the Lagrangian and the quantum origin of $\hbar$.

The paper is organised as follows: In Section 2 we start by giving some
mathematical considerations. In Section 3 we develop the fundamentals of
fourth-rank gravity. In Section 4 we consider the low energy regime and the
Schwarzschild solution. In Section 5 we apply fourth-rank gravity to
cosmology. Section 6 study the high-energy regime and the coupling to
conformal matter. Section 7 is dedicated to the conclusions. The Appendices
A, B and C, collect some standard results on Cosmography, General Relativity
and the Standard Model of Cosmology, respectively.

To our regret, due to the nature of this approach, in the Appendices we must
bore the reader by exhibiting some standard and well known results, but this
is necessary in order to illustrate where the new approach departs from the
standard one.

\section{2.}{Mathematical Preliminaries. Differentiable Manifolds}

Here we consider some elementary results for differentiable manifolds. Let us
start by considering the metric properties, which are related to the way in
which distances are measured. In what follows we take recourse to the
classical argumentation by Riemann.$^1$

Let $M$ be a {\it d}-dimensional differentiable manifold, and let $x^\mu$,
$\mu=0,\cdots,d-1$, be local coordinates. The infinitesimal element of
distance $ds$ should be a function of the coordinates $x$ and their
differentials $dx$'s

$$ds\,=\,f(x,\,dx)\,,\eqno{(2.1)}$$

\nt which is homogeneous of the first-order in $dx$'s

$$f(x,\,\lambda\,dx)\,=\,\lambda\,f(x,\,dx)\,,\eqno{(2.2a)}$$

\nt for $\lambda >0$, and is positive definite

$$f\,\geq\,0\,.\eqno{(2.2b1)}$$

Condition (2.2b1) was written in a time in which distances were, so to say,
positive. However, with the arrival of General Relativity one got used to
line elements with undefined signature. Condition (2.2b1) was there to
guarantee the invariance under the change $dx\rightarrow-dx$, {\it i.e.}, to
assure that distances measured when going in one direction are the same as
measured when going in the opposite direction. Therefore, we can replace
(2.2b1) by the weaker condition

$$f(x,\,-\,dx)\,=\,f(x,\,dx)\,.\eqno{(2.2b2)}$$

\nt Conditions (2.2a) and (2.2b2) can now we resumed into the single
condition

$$f(x,\,\lambda\,dx)\,=\,\mid\lambda\mid\,f(x,\,dx)\,,\eqno{(2.2)}$$

\nt with no restriction over the sign of $\lambda$.

Of course the possible solutions to (2.2) are infinitely many. Let us
restrict our considerations to monomial functions. Then we will have

$$ds\,=\,{(G_{\mu_1\cdots\mu_r}(x)\,dx^{\mu_1}\,\cdots\,dx^{\mu_r})}^{1\slash
r}\,.\eqno{(2.3)}$$

\nt In order for this quantity to be positive definite $r$ must be an even
number.

The simplest choice is $r=2$

$$ds^2\,=\,g_{\mu\nu}\,dx^\mu\,dx^\nu\,,\eqno{(2.4)}$$

\nt which corresponds to Riemannian geometry. The coefficients $g_{\mu\nu}$
are the components of the covariant metric tensor. The determinant of the
metric is defined by

$$g\,=\,{1\over d!}\,\epsilon^{\mu_1\cdots\mu_d}\,\epsilon^{\nu_1\cdots\nu_d}
\,g_{\mu_1\nu_1}\,\cdots\,g_{\mu_d\nu_d}\,.\eqno{(2.5)}$$

\nt If $g\not= 0$ we can define the inverse metric by

$$g^{\mu\nu}\,=\,{1\over(d-1)!}\,{1\over g}\,\epsilon^{\mu\mu_1\cdots\mu_{d-1
}}\,\epsilon^{\nu\nu_1\cdots\nu_{d-1}}\,g_{\mu_1\nu_1}\,\cdots\,g_{\mu_{d-1}
\nu_{d-1}}\,,\eqno{(2.6)}$$

\nt and satisfies

$$g^{\mu\lambda}\,g_{\lambda\nu}\,=\,\delta^\mu_\nu\,.\eqno{(2.7)}$$

\nt Densities of weight one can be constructed in terms of the quantity
$g^{1\slash2}$.

As pointed out by Riemann,$^1$ the next possibility is $r=4$. In this case
the line element is given by

$${ds}^4\,=\,G_{\mu\nu\lambda\rho}\,dx^\mu\,dx^\nu\,dx^\lambda\,dx^\rho\,.
\eqno{(2.8)}$$

\nt The coefficients $G_{\mu\nu\lambda\rho}$ are the components of a
covariant fourth-rank tensor. Since it is related to the metric properties of
the given manifold it is not an error to call it a "metric". The determinant
of the metric $G_{\mu\nu\lambda\rho}$ is defined as

$$G\,=\,{1\over d!}\,\epsilon^{\mu_1\cdots\mu_d}\,\cdots\,\epsilon^{\rho_1
\cdots\rho_d}\,G_{\mu_1\nu_1\lambda_1\rho_1}\,\cdots\,G_{\mu_d\nu_d\lambda_d
\rho_d}\,,\eqno{(2.9)}$$

\nt where the $\epsilon$'s can be chosen as the usual completely
antisymmetric Levi-Civita symbols. If $G\not= 0$ we can define the inverse
metric by

$$G^{\mu\nu\lambda\rho}\,=\,{1\over{(d-1)!}}\,{1\over G}\,\epsilon^{\mu\mu_1
\cdots\mu_{d-1}}\,\cdots\,\epsilon^{\rho\rho_1\cdots\rho_{d-1}}\,G_{\mu_1\nu_
1\lambda_1\rho_1}\,\cdots\,G_{\mu_{d-1}\nu_{d-1}\lambda_{d-1}\rho_{d-1}}\,.
\eqno{(2.10)}$$

\nt This inverse metric satisfies the relations

$$G^{\mu\alpha\beta\gamma}\,G_{\nu\alpha\beta\gamma}\,=\,\delta^\mu_\nu\,.
\eqno{(2.11)}$$

\nt That eq. (2.11) holds true for $G^{\mu\nu\lambda\rho}$ as defined in
(2.10) can be verified by hand in the two-dimensional case and with computer
algebraic manipulation for three and four dimensions.$^8$ Now, densities of
weight one can be constructed in terms of the quantity $G^{1\slash 4}$.

It is clear that fourth-rank geometry is observationally excluded at the
scale of distances of our daily life. However, a Riemannian behaviour can be
obtained for separable spaces. A space is said to be separable if
$G_{\mu\nu\lambda\rho}$ is of the form

$$G_{\mu\nu\lambda\rho}\,=\,g_{(\mu\nu}\,g_{\lambda\rho )}\,=\,{1\over3}\,(g_
{\mu\nu}\,g_{\lambda\rho}\,+\,g_{\mu\lambda}\,g_{\nu\rho}\,+\,g_{\mu\rho}\,g_
{\nu\lambda})\,.\eqno{(2.12)}$$

\nt In this case formula (2.8) reduces to (2.4). Separable metrics can also
be used as a quality control of later formal developments. In fact, all the
results and developments obtained for a generic metric
$G_{\mu\nu\lambda\rho}$ must reduce to those for Riemannian geometry when
applied to separable metrics.

In the case of a separable metric the determinant and the inverse metric are
given by

$$G\,=\,g^2\,,\eqno{(2.13a)}$$

$$G^{\mu\nu\lambda\rho}\,=\,{3\over{d+2}}\,g^{(\mu\nu}\,g^{\lambda\rho )}\,.
\eqno{(2.13b)}$$

Let us finish this Section with some considerations on the curvature
properties of manifolds. Curvature properties are described by the curvature
tensor

$$R^\lambda_{\rho\mu\nu}\,=\,\partial_\mu{\Gamma^\lambda}_{\nu\rho}\,-\,
\partial_\nu{\Gamma^\lambda}_{\mu\rho}\,+\,{\Gamma^\lambda}_{\mu\sigma}\,
{\Gamma^\sigma}_{\nu\rho}\,-\,{\Gamma^\lambda}_{\nu\sigma}\,{\Gamma^\sigma}_{
\mu\rho}\,,\eqno{(2.14)}$$

\nt constructed in terms of a connection ${\Gamma^\lambda}_{\mu\nu}$.

The metric and the connection are, in general, independent objects. They can
be related through a metricity condition. In Riemannian geometry the
metricity condition reads

$$\nabla_\lambda g_{\mu\nu}\,=\,\partial_\lambda g_{\mu\nu}\,-\,{\Gamma^\rho}
_{\lambda\mu}\,g_{\rho\nu}\,-\,{\Gamma^\rho}_{\lambda\nu}\,g_{\mu\rho}\,=\,0
\,.\eqno{(2.15)}$$

\nt The number of unknowns for a symmetric connection  ${\Gamma^\lambda}_{\mu
\nu}$ and the number of equations (2.15) are the  same,  ${1\over2}d^2(d+1)$.
Therefore, since this is an algebraic linear system, the solution  is  unique
and is given by the familiar Christoffel symbols of the second kind

$${\Gamma^\lambda}_{\mu\nu}\,=\,\lbrace^\lambda_{\mu\nu}\rbrace(g)\,=\,{1
\over2}\,g^{\lambda\rho}\,(\partial_\mu g_{\nu\rho}\,+\,\partial_\nu g_{\mu
\rho}\,-\,\partial_\rho g_{\mu\nu})\,.\eqno{(2.16)}$$

\nt Therefore, in Riemannian geometry one can talk of the curvature
properties of a metric $g_{\mu\nu}$. This can be done because there exist a
natural connection, the Christoffel symbol of the second kind, in terms of
which we can construct a curvature tensor.

In the case of a fourth-rank metric a condition analogous to (2.15) would
read

$$\nabla_\mu G_{\alpha\beta\gamma\delta}\,=\,\partial_\mu G_{\alpha\beta
\gamma\delta}\,-\,{\Gamma^\nu}_{\mu\alpha}\,G_{\nu\beta\gamma\delta}\,-\,{
\Gamma^\nu}_{\mu\beta}\,G_{\alpha\nu\gamma\delta}\,-\,{\Gamma^\nu}_{\mu
\gamma} \,G_{\alpha\beta\nu\delta}\,-\,{\Gamma^\nu}_{\mu\delta}\,G_{\alpha
\beta\gamma\nu}\,=0\,.\eqno{(2.17)}$$

\nt However, in this case, the number of unknowns ${\Gamma^\lambda}_{\mu\nu}$
is, as before, ${1\over2}d^2(d+1)$, while the number of equations is

$${1\over24}\,d^2\,(d\,+\,1)\,(d\,+\,2)\,(d\,+\,3)\,>\,{1\over2}\,d^2\,(d\,+
\,1)\,.\eqno{(2.18)}$$

\nt Therefore the system is overdetermined and some differentio-algebraic
conditions must be satisfied by the metric. Since, in general, such
restrictions will not be satisfied by a generic metric, one must deal with
${\Gamma^\lambda}_{\mu\nu}$ and $G_{\mu\nu\lambda\rho}$ as independent
objects. Therefore, for physical applications, the connection and the metric
must be considered as independent fields.

A metricity condition can be imposed consistently only if the number of
independent components of the metric is less than that naively implied by
(2.18). The maximum acceptable number of independent components is ${1\over2}
d(d+1)$. This can be achieved, for instance, if the metric is a separable
one. Furthermore, one can verify that in this case the metricity condition
(2.17) reduces to the usual metricity condition (2.15) for the metric
$g_{\mu\nu}$ and therefore ${\Gamma^\lambda}_{\mu\nu}$ is precisely that for
Riemannian geometry, i.e., the Christoffel symbol of the second kind.

\section{3.}{Conformal Fourth-Rank Gravity}

In this Section we develop a theory for the gravitational field based on
fourth-rank geometry. The use of fourth-rank geometry is motivated by the
following considerations. At very high energies the masses of particles
involved in physical processes become negligible as compared to the energies,
in fact they can be set equal to zero. Therefore, there is no fundamental
mass setting the scale of energies, and all physical processes must be scale
invariant. This especulation is confirmed by experiments, such as deep
inelastic scattering, which show that, in fact, at very high energies,
physical processes are scale invariant. It can furthermore be shown, from a
mathematical point of view, that scale invariance is equivalent to conformal
invariance. Therefore, high-energy physics is associated to a geometry
exhibiting, in a model independent way, conformal invariance in 4 dimensions.
In another work$^4$ we show that the critical dimension, for which field
theories are integrable, is equal to the rank of the metric. Therefore, if we
want to construct a field theory in 4 dimensions showing agreement with the
observed conformal invariance at high energies we must take recourse to
fourth-rank geometry.

We  arrive  therefore  to  the  following   scheme:   at   short   distances,
high-energies, the geometry is of fourth-rank while at large distances, low
energies, the geometry is of second-rank, Riemannian. It is clear furthermore
that the Riemannian behaviour of  the  geometry  must  be  recovered  as  the
low-energy limit of the high-energy theory. This  would  be  possible  if  at
low-energies the fourth-rank metric tensor $G_{\mu\nu\lambda\rho}$ becomes
separable. In this case the line element factors and one is back to the
Riemannian case. This would explain why the universe, even when described by
a fourth-rank metric, looks Riemannian at large, low energy, scales. The
problem is now to obtain this Riemannian behaviour as the low-energy regime
of some field theory.

The conformal invariance requirement determines, almost uniquely the
geometrical invariant to be used as Lagrangian. The field equations exhibit
three energy regimes: low, medium, and high. In the low-energy regime there
is no matter and the fourth-rank metric is separable, $G_{\mu\nu\lambda\rho}=
g_{(\mu\nu}g_{\lambda\rho )}$. Then the line element factors and one is back
to the Riemannian case. In the medium-energy regime the geometry is still
Riemannian, $G_{\mu\nu\lambda\rho}=g_{(\mu\nu}g_{\lambda\rho )}$, but there
is matter involved in the game. In this case the gravitational field couples
in a different way, as compared to General Relativity, to matter. Lastly, we
have the true high-energy regime in which there is matter and the geometry is
truly fourth-rank. These energy regimes, and their observational
consequences, are further analysed in Section 4, 5, and 6.

\subsection{3.1}{Fourth-Rank Gravitational Equations}

As in General Relativity, in order to describe the dynamics of the
gravitational field we need to construct a geometrical invariant. From the
metric alone it is impossible to construct any invariant, apart from the
trivial solution: a constant. Therefore we must take recourse to another
geometrical object. The necessary object is the Ricci tensor for an arbitrary
connection ${\Gamma^\lambda}_{\mu\nu}$, which is obtained as a contraction of
the Riemann tensor, defined in (2.14),

$$R_{\mu\nu}\,=\,R^\lambda_{\mu\lambda\nu}\,=\,\partial_\lambda{\Gamma^
\lambda}_{\mu\nu}\,-\,\partial_\nu{\Gamma^\lambda}_{\lambda\nu}\,+\,{\Gamma^
\lambda}_{\lambda\sigma}\,{\Gamma^\sigma}_{\mu\nu}\,-\,{\Gamma^\lambda}_{\mu
\sigma}\,{\Gamma^\sigma}_{\lambda\nu}\,.\eqno{(3.1)}$$

\nt The simplest invariants which can be constructed with the metric
$G_{\mu\nu\lambda\rho}$ and the Ricci tensor $R_{\mu\nu}$ are

$$\langle R^2\rangle\,=\,G^{\mu\nu\lambda\rho}\,R_{\mu\nu}\,R_{\lambda\rho}
\,,$$

$$\langle R^4\rangle\,=\,G^{\mu\nu\lambda\rho}\,G^{\alpha\beta\gamma\delta}\,
R_{\mu\alpha}\,R_{\nu\beta}\,R_{\lambda\gamma}\,R_{\rho\delta}\,,\quad etc.
\eqno{(3.2)}$$

\nt The Lagrangian therefore will be of the form

$${\cal L}\,=\,L(\langle R^2\rangle,\,\langle R^4\rangle,\,\cdots)\,G^{1
\slash4}\,,\eqno{(3.3)}$$

\nt where $G$ is the determinant of $G_{\mu\nu\lambda\rho}$. The scalar
function $L$ to be put in (3.3) should make the Lagrangian a conformally
invariant function. Under rescalings of the metric

$$G_{\mu\nu\lambda\rho}\,\rightarrow\,\lambda\,G_{\mu\nu\lambda\rho}\,,
\eqno{(3.4)}$$

\nt the inverse metric $G^{\mu\nu\lambda\rho}$ and $G^{1\slash4}$ transform
as

$$G^{\mu\nu\lambda\rho}\,\rightarrow\,\lambda^{-1}\,G^{\mu\nu\lambda\rho}\,,
\eqno{(3.5a)}$$

$$G^{1\slash4}\,\rightarrow\,\lambda\,G^{1\slash 4}\,.\eqno{(3.5b)}$$

\nt Therefore the Lagrangian should be of the form

$${\cal L}\,=\,[\alpha\,\langle R^2\rangle\,+\,\beta\,{{\langle R^4\rangle}
\over{\langle R^2\rangle}}\,+\,\cdots]\,G^{1\slash4}\,.\eqno{(3.6)}$$

\nt However, all the terms after the first one, are highly non-local.
Therefore, the only sensible solution is

$${\cal L}_{CG}\,=\,\kappa_{CG}\,\langle R^2\rangle\,G^{1\slash4}\,,
\eqno{(3.7)}$$

\nt where the coupling constant

$$\kappa_{CG}\,\approx\,\kappa_E\,{L_{Planck}}^2\,=\,{{\hbar c}\over{8\pi}}
\,,\eqno{(3.8)}$$

\nt is the Einstein gravitational constant $\kappa_E={c^4\over{8\pi G}}$,
times a constant of the order of ${L_{Planck}}^2$.

The above is the analogue of the Palatini Lagrangian for General Relativity.
But now, since there is no metricity condition, a Lagrangian analogous to the
Einstein-Hilbert one simply does not exist.

The total Lagrangian must consider also the contributions of matter and is
given by

$${\cal L}\,=\,{\cal L}_{CG}\,+\,{\cal L}_{matter}\,.\eqno{(3.9)}$$

\nt Variation with respect to the connection gives

$${{\delta{\cal L}}\over{\delta{\Gamma^\lambda}_{\mu\nu}}}\,=\,{{\delta{\cal
L}_{CG}}\over{\delta{\Gamma^\lambda}_{\mu\nu}}}\,+\,{{\delta{\cal L}_{matter}
}\over{\delta{\Gamma^\lambda}_{\mu\nu}}}\,=\,0\,,\eqno{(3.10)}$$

\nt where

$${{\delta{\cal L}_{CG}}\over{\delta{\Gamma^\lambda}_{\mu\nu}}}\,=\,{{
\partial{\cal L}_{CG}}\over{\delta{\Gamma^\lambda}_{\mu\nu}}}\,-\,d_\rho\left
({{\partial{\cal L}_{CG}}\over{\partial(\partial_\rho{\Gamma^\lambda}_{\mu\nu
}}}\right)$$

$$=\,\gamma^{\alpha\beta}\,[{1\over2}\,(\delta^\nu_\lambda\,{\Gamma^\mu}_{
\alpha\beta}\,+\,\delta^\mu_\lambda\,{\Gamma^\nu}_{\alpha\beta})\,+\,\delta^
\mu_\alpha\,\delta^\nu_\beta\,{\Gamma^\sigma}_{\lambda\sigma}\,-\,\delta^\nu_
\beta\,{\Gamma^\mu}_{\lambda\alpha}\,-\,\delta^\mu_\beta\,{\Gamma^\nu}_{
\lambda\alpha}]\,G^{1\slash4}$$

$$-\,d_\rho[\gamma^{\alpha\beta}\,\left(\delta^\rho_\lambda\,\delta^\mu_\beta
\,\delta^\nu\alpha\,-\,{1\over2}\,\delta^\rho_\beta\,(\delta^\mu_\lambda\,
\delta^\nu_\alpha\,+\,\delta^\nu_\lambda\,\delta^\mu_\alpha)\right)\,G^{1
\slash4}]\,.\eqno{(3.11)}$$

\nt with

$$\gamma^{\alpha\beta}\,=\,G^{\alpha\beta\gamma\delta}\,R_{\gamma\delta}\,,
\eqno{(3.12)}$$

\nt (for simplicity, we have omitted $\kappa_{CG}$). In all known cases of
physical interest the matter Lagrangian does not depend on the
connection.$^5$ Therefore the second term in (3.10) vanishes and one remains
with a metricity condition which has the solution

$${\Gamma^\lambda}_{\mu\nu}\,=\,\lbrace^\lambda_{\mu\nu}\rbrace(\gamma)\,=\,
{1\over2}\,\gamma^{\lambda\rho}\,(\partial_\mu\gamma_{\nu\rho}\,+\,\partial_
\nu\gamma_{\nu\rho}\,-\,\partial_\rho\gamma_{\mu\nu})\,,\eqno{(3.13)}$$

\nt i.e., the connection is the Christoffel symbol of the second kind for the
tensor $\gamma_{\mu\nu}$, which we have assumed to be regular. We can
therefore write

$$R_{\mu\nu}(\Gamma)\,=\,R_{\mu\nu}(\gamma)\,.\eqno{(3.14)}$$

\nt Furthermore

$$\langle R^2\rangle\,=\,G^{\mu\nu\lambda\rho}\,R_{\mu\nu}\,R_{\lambda\rho}\,
=\,\gamma^{\mu\nu}\,R_{\mu\nu}(\gamma)\,=\,R(\gamma)\,.\eqno{(3.15)}$$

Variation with respect to $G_{\mu\nu\lambda\rho}$

$${{\delta{\cal L}}\over{\delta G^{\mu\nu\lambda\rho}}}\,=\,{{\partial{\cal L
}}\over{\partial G^{\mu\nu\lambda\rho}}}\,=\,d_\sigma\left({{\partial{\cal L}
}\over{\partial (\partial_\sigma G^{\mu\nu\lambda\rho})}}\right)\,=\,0\,,
\eqno{(3.16)}$$

\nt gives

$$\kappa_{CG}\,[ R_{(\mu\nu}\,R_{\lambda\rho )}\,-\,{1\over4}\,\langle R^2
\rangle\,G_{\mu\nu\lambda\rho}]\,=\,T_{\mu\nu\lambda\rho}\,.\eqno{(3.17)}$$

\nt where

$$T_{\mu\nu\lambda\rho}\,=\,-\,G^{-1\slash4}\,{{\delta{\cal L}_{matter}}\over
{\delta G^{\mu\nu\lambda\rho}}}\,,\eqno{(3.18)}$$

\nt is the fourth-rank energy-momentum tensor.

More information can be obtained from eq. (3.17) by observing that the
energy-momentum tensor must decompose into one part proportional to the
metric and another part which is a separable tensor. In order to accommodate
all the symmetries is necessary to have

$$T_{\mu\nu\lambda\rho}\,=\,{{{L_{Planck}}^4}\over{\kappa_{CG}}}\,[S_{4,(\mu
\nu}\,S_{4,\lambda\rho)}\,-\,{1\over4}\,\langle{S_4}^2\rangle\,G_{\mu\nu
\lambda\rho}]\,,\eqno{(3.19)}$$

\nt where

$$\langle{S_4}^2\rangle\,=\,G^{\mu\nu\lambda\rho}\,S_{4,\mu\nu}\,S_{4,\lambda
\rho}\,.\eqno{(3.20)}$$

\nt In this case the field equations reduce to the simple form

$$\kappa_E\,R_{\mu\nu}(\gamma)\,=\,\pm\,S_{4,\mu\nu}\,;\eqno{(3.21)}$$

\nt and, as a further consequence we have

$$\kappa_E\,R(\gamma)\,=\,\kappa_E\,\langle R^2\rangle\,=\,\langle{S_4}^2
\rangle\,=\,{S_4}^2(\gamma)\,,\eqno{(3.22)}$$

\nt where $S_4(\gamma )=\gamma^{\mu\nu}S_{4,\mu\nu}(\gamma )$. One would be
tempted to replace $S_{4,\mu\nu}$ by the reduced energy-momentum tensor
appearing in (B.14). However, that tensor is derived from a Lagrangian
containing a metric $g_{\mu\nu}$, an object which is, in principle, absent in
fourth-rank geometry. Concerning the $\pm$ sign in (3.21), this must be
determined by taking recourse to some application, as will be done in Section
4.

\subsection{3.2}{The Different Energy Regimes}

The field equations (3.17) exhibit three energy regimes: low, medium and
high. In the low-energy regime there is no matter and therefore the geometry
is Riemannian, $G_{\mu\nu\lambda\rho}=g_{(\mu\nu}g_{\lambda\rho)}$, as can
be read from (3.17). In this case the field equations do not reduce to the
Einstein field equations in vacuum. In the medium-energy regime the geometry
is still Riemannian, $G_{\mu\nu\lambda\rho}=g_{(\mu\nu}g_{\lambda\rho)}$, but
now there is matter in the game. This possibility is not excluded as a closer
analysis of eqs. (3.17) reveals. Finally, we have the true high-energy regime
in which there is matter and the geometry is truly fourth-rank.

\subsubsection{3.2.1.}{The Low-Energy Regime}

In the low-energy regime ${\cal L}_{matter}=0$ and then the field equations
reduce to

$$R_{(\mu\nu}\,R_{\lambda\rho )}\,-\,{1\over4}\,\langle R^2\rangle\,G_{\mu\nu
\lambda\rho}\,=\,0\,.\eqno{(3.23)}$$

\nt The only sensible solution is

$$G_{\mu\nu\lambda\rho}\,=\,g_{(\mu\nu}\,g_{\lambda\rho)}\,,\eqno{(3.24a)}$$

$$R_{\mu\nu}\,=\,{1\over2}\,{\langle R^2\rangle}^{1\slash2}\,g_{\mu\nu}\,,
\eqno{(3.24b)}$$

\nt and therefore the geometry is Riemannian.

The tensor $\gamma^{\mu\nu}$ is given by

$$\gamma^{\mu\nu}\,=\,{1\over2}\,{\langle R^2\rangle}^{1\slash2}\,g^{\mu\nu}
\,,\eqno{(3.25)}$$

$$\gamma_{\mu\nu}\,=\,2\,{\langle R^2\rangle}^{-1\slash2}\,g_{\mu\nu}\,=\,2\,
R^{-1\slash2}(\gamma)\,g_{\mu\nu}\,.\eqno{(3.26)}$$

\nt Then, eq. (3.24b) is rewritten as

$$R_{\mu\nu}(\gamma)\,-\,{1\over4}\,R(\gamma)\,g_{\mu\nu}=0\,.\eqno{(3.27)}$$

\nt One must therefore compute equations (3.27) for a tensor $\gamma_{\mu
\nu}$, and then the physical metric $g_{\mu\nu}$ is obtained from (3.26).

Let us furthermore observe that the dimensions of $\gamma^{\mu\nu}$,
$\gamma_{\mu\nu}$, $R_{\mu\nu}(\gamma g)$ and $R(\gamma )$ are given by

$$dim(\gamma^{\mu\nu})\,=\,L^{-2}\,,$$

$$dim(\gamma_{\mu\nu})\,=\,L^2\,,$$

$$dim(R_{\mu\nu}(\gamma))\,=\,L^{-2}\, ,$$

$$dim(R(\gamma))\,=\,L^{-4}\,.\eqno{(3.28)}$$

Let us now rewrite the field equations (3.27) in terms of the metric
$g_{\mu\nu}$. Let us start by rewriting eq. (3.26) as

$$\gamma_{\mu\nu}\,=\,\lambda^2\,e^\psi\,g_{\mu\nu}\,,\eqno{(3.29)}$$

\nt where

$$\lambda^2\,e^\psi\,=\,2\,R^{-1\slash2}(\gamma)\,,\eqno{(3.30)}$$

\nt and $\lambda$ has dimensions of length. Therefore

$$R(\gamma)\,=\,{4\over{\lambda^{-4}}}\,e^{-2\psi}\,.\eqno{(3.31)}$$

The Ricci tensors are related by$^9$

$$R_{\mu\nu}(\gamma)\,=\,R_{\mu\nu}(g)\,+\,\nabla_\mu\psi_\nu\,-\,{1\over2}\,
\psi_\mu\,\psi_\nu\,+\,{1\over2}\,g_{\mu\nu}\,({\nabla_g}^2\psi\,+\,g^{
\alpha\beta}\,\psi_\alpha\,\psi_\beta)\,,\eqno{(3.32)}$$

\nt while the scalar curvatures are related by

$$R(\gamma)\,=\,{1\over{\lambda^{-2}}}\,e^{-\psi}\,[ R(g)\,+\,3\,{\nabla_g}^2
\psi\,+\,{3\over2}\,g^{\alpha\beta}\,\psi_\alpha\,\psi_\beta]\,.
\eqno{(3.33)}$$

\nt The field equations are rewritten as

$$R_{\mu\nu}(g)\,+\,\nabla_\mu\psi_\nu\,-\,{1\over2}\,\psi_\mu\,\psi_\nu\,-\,
{1\over4}\, g_{\mu\nu}\,[ R(g)\,+\,{\nabla_g}^2\psi\,-\,{1\over2}\,g^{\alpha
\beta}\,\psi_\alpha\,\psi_\beta]\,=\,0\,.\eqno{(3.34)}$$

Combining (3.31) and (3.33) we obtain the differential equation for the
conformal factor $\psi$

$$e^{-\psi}\,[ R(g)\,+\,3\,{\nabla_g}^2\psi\,+\,{3\over2}\,g^{\alpha\beta}\,
\psi_\alpha\,\psi_\beta]\,=\,{4\over{\lambda^{-2}}}\,e^{-2\psi}\,.
\eqno{(3.35)}$$

As mentioned in the introduction, in vacuum, General Relativity is in
excelent agreement with observation. Therefore, in this regime, our theory
must coincide with General Relativity. This is not evident from eqs. (3.27)
and in fact they are not equivalent. Therefore, the equivalence must be
established at the level of the solutions rather than of the field equations.
This regime is further explored in Section 4.

\subsubsection{3.2.2.}{The Medium-Energy Regime}

In the medium-energy regime the metrics $G_{\mu\nu\lambda\rho}$ and
$g_{\mu\nu}$ are still related by (3.24a). In this case it is therefore
reasonable to replace $S_{4,\mu\nu}$ with that appearing in (B.14)

$$\kappa_E\,R_{\mu\nu}(\gamma)\, =\,\pm\,S_{2,\mu\nu}(g)\,.\eqno{(3.36)}$$

\nt However, the field equations (3.36) are not equivalent to Einstein field
equations since the Ricci tensor appearing here is for the tensor
$\gamma_{\mu\nu}$ and not for the metric $g_{\mu\nu}$. The above choice is a
delicate point since other mechanisms of coupling the fourth-rank geometry
with "second-rank" matter can be conceived. For example one can consider

$$\kappa_E\,G_{\mu\nu}(\gamma)\,=\,\pm\,T_{2,\mu\nu}(g)\,,\eqno{(3.37)}$$

\nt which is not equivalent to (3.36). We have tested this and other
possibilities and we have concluded that (3.36) is the correct choice. Which
sign is to be chosen in (3.36) must be decided by considering some
application.

The large scale geometric structure of our universe seems to be well
described by Riemannian geometry, and since matter is involved, its
description belongs to medium-energy regime. We explore this possibility in
Section 5.

\subsubsection{3.2.3.}{The High-Energy Regime}

In this case $G_{\mu\nu\lambda\rho}$ is not a separable metric. Furthermore,
the energy-momentum tensor of matter is traceless, a property which is
equivalent to scale invariance. Therefore, scale invariance is present at
very high-energies, and one can confidently consider this as the high-energy
regime of the theory. This regime is further explored in Section 6.

\section{4.}{The Low-Energy Regime}

Now we explore the consequences of our field equations in the absence of
matter. Our field equations do not reduce to the vacuum Einstein field
equations. Therefore, the observational equivalence must be established at
the level of the solutions rather than at the level of the field equations.

\subsection{4.1.}{Static Spherically Symmetric Fields}

In order to check the validity of the field equations (3.27) we consider the
standard test of a spherically symmetric field. The line element is given by

$${ds}^2\,=\,A(r)\,{dt}^2\,-\,B(r)\,{dr}^2\,-\,r^2\,{d\Omega}^2\,,
\eqno{(4.1)}$$

\nt where

$${d\Omega}^2\,=\,{d\theta}^2\,+\,{sin}^2\theta\,{d\varphi}^2\,.
\eqno{(4.2)}$$

\nt One can certainly assume that $R(\gamma)$ will be a function of $r$ only.
The physical metric $g_{\mu\nu}$ and the tensor $\gamma_{\mu\nu}$ are related
by

$$\gamma_{\mu\nu}\,=\,f(R(\gamma))\,g_{\mu\nu}\,=\,f(r)\,g_{\mu\nu}\,.
\eqno{(4.3)}$$

\nt Therefore, the line element associated to the tensor $\gamma_{\mu\nu}$ is

$${ds_\gamma}^2\,=\,f(r)\,[ A(r)\,{dt}^2\,-\,B(r)\,{dr}^2\,-\,r^2\,{d\Omega}
^2]\,,\eqno{(4.4)}$$

\nt and, by a redefinition of $r$, it can be rewritten as in (4.1)

$${ds_\gamma}^2\,=\,\lambda^2\,[{\bar A}(r)\,{dt}^2\,-\,{\bar B}(r)\,{dr}^2\,
-\, r^2\,{d\Omega}^2]\,,\eqno{(4.5)}$$

\nt where $\lambda$ has dimensions of length.

The solution is the Kottler metric$^6$

$$\gamma_{00}\,=\,\lambda^2\,(a\,+\,{b\over r}\,+\,c\,r^2)\,,$$

$$\gamma_{11}\,=\,-\,\lambda^2\,{(a+{b\over r}+c\,r^2)}^{-1}\,.\eqno{(4.6)}$$

\nt The associated scalar curvature $R(\gamma)$ is constant

$$R(\gamma)\,=\,12\,{c\over a}\,\lambda^{-2}\,.\eqno{(4.7)}$$

We can now write the field equations (3.27) in terms of the metric
$g_{\mu\nu}$

$$R_{\mu\nu}(g)\,-\,{1\over4}\,R(g)\,g_{\mu\nu}\,=\,0\,.\eqno{(4.8)}$$

\nt This is possible due to the fact that the Ricci tensor is homogeneous of
order zero in $g_{\mu\nu}$. Therefore, the constant conformal factor,
essentially (4.7), will cancel from the field equations. Then, the metric
$g_{\mu\nu}$ will be the Kottler metric as given in (4.6)

$$g_{00}\,=\,\left(a\,+\,{b\over r}\,+\,c\,r^2\right)\,,$$

$$g_{11}\,=\,-\,{\left(a\,+\,{b\over r}\,+\,c\,r^2\right)}^{-1}\,,
\eqno{(4.9)}$$

\nt Now we can put $c=0$ and obtain the Schwarzschild metric.

This  long  detour  was  necessary  in  order  to  check   that   the   limit
$c\rightarrow0$ was a consistent procedure.

\subsection{4.2.}{Comments}

This is the weakest energy regime and coincides with General Relativity.
Therefore, the Schwarzschild solution, the Newtonian limit and the properties
of gravitational radiation will be the same as for General Relativity. One
must therefore not check that the proper limit exists but that the
observables departures agree with observation. For this we must turn our
attention to the two following regimes.

\section{5.}{The Medium-Energy Regime. Fourth-Rank Cosmology}

Now we explore the consequences of our field equations in the medium-energy
regime. The ideal laboratory is the universe. The large scale geometry of the
universe seems to be Riemannian. Furthermore there is matter present.
Therefore, in the context of fourth-rank gravity, the description of the
universe belongs to the medium-energy regime. The metric
$G_{\mu\nu\lambda\rho}$ will be separable in terms of a metric $g_{\mu\nu}$
which we assume to be the FRW metric. Matter is described by a perfect fluid;
therefore we use the energy-momentum tensor appearing in (A.7).

When fourth-rank gravity is applied to cosmology one should deal with
equations analogous to the Einstein-Friedman equations of the Standard Model
of Cosmology. In fourth-rank gravity however, matter enters the field
equations in a non-linear way. An essential difference with respect to
General Relativity is the fact that the equations determining the evolution
of the universe involve not only the energy density and the pressure but also
their time derivatives. Therefore, in order to correctly deal with these
equations, one should provide a time dependent state equation. As a first
approach we restrict our considerations to the case of a time independent
state equation. Of course, this is a quite strong assumption.

The theory predicts an increasing total entropy such that the expansion of
the universe is an adiabatic non-isoentropic process. Therefore, the
evolution of the universe, in the framework of fourth-rank cosmology is, as
expected on physical grounds, an irreversible process.

The following conclusions are obtained after incorporating $k_{obs}=0$.

For the early universe matter is described by a state equation in which $y
\approx{1\over3}$. In this case $q>0$ and from this fact one deduces the
existence of a very dense state of matter at some time in the past. Causality
is not violated for $t>t_{class}\approx{10}^{19}t_{Planck}\approx{10}^{-24}
s$. At earlier times quantum mechanical effects dominate the scene. In  fact,
the radius of the universe is exactly the Compton  wavelength  associated  to
its mass. Our classical approach breaks down so  that  the  very  concept  of
causality is meaningless. Therefore, there is no violation of  causality,  or
horizon problem.

For the present Universe it is necessary to assume $q<0$; this does not
contradict the observed expansion of the universe from an initial hot ball.
In General Relativity $q>0$ and $a(t)$ is a convex function of t. Due to this
fact one is used to think of the evolution of the universe with $q>0$.
However, an evolution from an initial singularity may also be conceived with
a concave function, $q<0$. The field equations predict $\Omega\approx4y$,
where $y={p\over\rho}$. For the present universe we use $\Omega_{small}=0.01$
and we obtain $y_{pred}=2.5\times{10}^{-3}$. {\it y} can be estimated from
the mean random velocity of typical galaxies to be
$y_{random}=1\times{10}^{-5}$.

\vfill\eject

\subsection{5.1.}{The Field Equations}

The field equations are

$$\kappa_E\,R_{\mu\nu}(\gamma)\,=\,\pm\,[(\rho\,+\,p)\,u_\mu\,u_\nu\,-\,{1
\over2}\,(\rho\,-\,p)\,g_{\mu\nu}]\,.\eqno{(5.1)}$$

\nt On the other hand we have

$$\gamma^{\mu\nu}\,=\,G^{\mu\nu\lambda\rho}\,R_{\lambda\rho}\,=\,{1\over{3
\kappa_E}}\,[(\rho\,+\,p)\,u^\mu\,u^\nu\,-\,(\rho\,-\,2\,p)\,g^{\mu\nu}]\,.
\eqno{(5.2)}$$

The inverse of (5.2) is given by

$$\gamma_{\mu\nu}\,=\,3\,\kappa_E\,[{(\rho+p)\over{3p(\rho-2p)}}\,u_\mu\,u_
\nu\,-\,{1\over{(\rho-2p)}}\,g_{\mu\nu}]\,.\eqno{(5.3)}$$

\nt There is a global $\pm$ sign in (5.2) and (5.3) which we have fixed at
will since the final result is independent of this choice.

The first step is to calculate the Ricci tensor for the metric
$\gamma_{\mu\nu}$. Let us start by writing the associated line element

$${ds_\gamma}^2\,=\,{1\over p}\,{dt}^2\,+\,{3\over{(\rho-2p)}}\,a^2\,{d\ell}
^2\,.\eqno{(5.4)}$$

\nt We have omitted the constant in front of (5.3) since the final result is
independent of this factor. In what follows we assume $p>0$ and $\rho
-2p>0$. We assume furthermore that $\rho$ and $p$ are functions of $t$
only. Then we can introduce the new time coordinate

$$d\tau\,=\,{1\over{p^{1\slash2}}}\,dt\,.\eqno{(5.5)}$$

\nt Then the line element (5.5) is rewritten as

$${ds}^2\,=\,{d\tau}^2\,+\,A^2\,{d\ell}^2\,,\eqno{(5.6)}$$

\nt with

$$A\,=\,[{({3\over{(\rho-2p)}})}^{1\slash2}\,a](\tau)\,.\eqno{(5.7)}$$

\nt The above is nothing more than a FRW line element with Euclidean
signature. We can therefore use eqs. (A.3) with $a^2\rightarrow-A^2$. The
Ricci tensor is then given by

$$R_{\mu\nu}(\gamma)\,=\,-\,{2\over{A^2}}\,[A\,A''\,-\,(-\,k\,+\,{A'}^2)]\,u_
\mu\,u_\nu\,-\,{1\over{A^2}}\,[A\,A''\,+\,2\,(-\,k\,+\,{A'}^2)]\,\gamma_{\mu
\nu}\,,\eqno{(5.8)}$$

\nt where primes denote derivatives with respect to $\tau$. In the system of
coordinates involving $t$ the above expression is given by

$$R_{\mu\nu}(\gamma)\,=\,-\,{2\over{pA^2}}\,[A\,A''\,-\,(-\,k\,+\,{A'}^2)]\,
\delta^0_\mu\,\delta^0_\nu\,-\,{1\over{A^2}}\,[A\,A''\,+\,2\,(-\,k\,+\,{A'}^2
)]\,\gamma_{\mu\nu}\,,\eqno{(5.9)}$$

Comparison with the Ricci tensor obtained from the field equations, eq.
(5.1), gives

$$-\,3\,\kappa_E\,{1\over p}\,{A''\over A}\,=\,\pm\,{1\over2}\,(1\,+\,3\,y)\,
\rho\,,\eqno{(5.10a)}$$

$$-\,3\,\kappa_E\,{1\over{(\rho-2p)}}\,{1\over{A^2}}\,[ A\,A''\,+\,2\,(-\,k\,
+\,{A'}^2)]\,=\,\pm\,{1\over2}\,(1\,-\,y)\,\rho\,.\eqno{(5.10b)}$$

\nt For the applications the field equations are better rewritten as

$$6\,\kappa_E\,{1\over{(\rho -2p)}}\,{1\over{A^2}}\,(-\,k\,+\,{A'}^2)\,=\,
\mp\,{1\over2}\,{{(1-4y-y^2)}\over{(1-2y)}}\,\rho\,,\eqno{(5.11a)}$$

$$-\,{{(1-4y-y^2)}\over{(1-2y)}}\,{1\over p}\,A\,A''\,+\,2\,(1\,+\,3\,y)\,
{1\over{(\rho-2p)}}\,(-\,k\,+\,{A'}^2)\,=\,0\,.\eqno{(5.11b)}$$

\nt The field equations written in this form are of practical use since the
first one allows us to determine the value of $k$ when evaluated at the
present time. The second one allows us to determine the evolution of the
early universe.

\subsection{5.2.}{The Entropy of the Universe}

The entropy variation is governed by

$$T\,dS\,=\,dE\,+\,p\,dV\,=\,d(r\,a^3)\,+\,p\,d(a^3)$$

$$=\,{{(1+3y)(1-y^2)}\over{(1-4y-y^2)}}\,\rho\,a^2\,da\,+\,\rho\,a^3\,{{2(1+8
y^2+16y^3-5y^4)}\over{(1-2y)(1-2y+5y^2)(1-4y-y^2)}}\,dy\,.\eqno{(5.12)}$$

\nt Since the radius of the universe grows at a rate much larger than that by
which y decreases, the above quantity is positive. Hence the theory predicts,
in a natural  way,  an  increasing  total  entropy  of  the  universe.  Thus,
fourth-rank cosmology predicts an adiabatic non-isoentropic, and therefore
irreversible, expansion of the universe.

The next step is to go back to the time coordinate $t$. This contains the
time dependence of $p$ on $t$. We assume that the almost pressureless regime
has lasted for such a long time that we can confidently work under the
assumption that $p$ and $\rho$ are constant, i.e., a time independent state
equation. This is done now.

\vfill\eject

\subsection{5.3.}{Constant y}

We assume that the almost pressureless regime has lasted for such a long time
that we can confidently work under the assumption that $p$ and $\rho$ are
constant, i.e., a time independent state equation. In this case the relevant
equations are obtained with the simple replacements

$$A\,\rightarrow{\left({3\over{(\rho-2p)}}\right)}^{1\slash2}\,a\,,
\eqno{(5.13a)}$$

$$(\,)'\,\rightarrow\,p^{1\slash2}\,(\,)^.\,.\eqno{(5.13b)}$$

\nt In this case eq. (5.10a) is rewritten like

$$-\,3\,\kappa_E\,{{\ddot a}\over a}\,=\,\pm\,{1\over2}\,(1\,+\,3\,y)\,\rho\,
.\eqno{(5.14)}$$

\nt Equations (5.11) reduce to

$$6\,\kappa_E\,{1\over{a^2}}\,(-\,k\,+\,{3y\over{(1-2y)}}\,{\dot a}^2)\,=\,
\mp\,{3\over2}\,{{(1-4y-y^2)}\over{(1-2y)}}\,\rho\,,\eqno{(5.15a)}$$

$$-\,3\,{{(1-4y-y^2)}\over{(1-2y)}}a\,{\ddot a}\,+\,2\,(1\,+\,3\,y)\,
(-\,k\,+\,{3y\over{(1-2y)}}\,{\dot a}^2)\,=\,0\,.\eqno{(5.15b)}$$

\subsection{5.4.}{Incorporating Flatness}

Let us now incorporate the observed fact $k_{obs}=0$. Then, eqs. (5.15)
reduce to

$$18\,\kappa_E\,{y\over{(1-2y)}}\,{{\dot a}^2/over{a^2}}\,=\,\mp\,{3\over2}\,
{{(1-4y-y^2)}\over{(1-2y)}}\,\rho\,,\eqno{(5.16a)}$$

$$-\,3\,{{(1-4y-y^2)}\over{(1-2y)}}a\,{\ddot a}\,+\,6\,{y(1+3y)\over{(1-2y)}}
\,{\dot a}^2\,=\,0\,.\eqno{(5.16b)}$$

For the physically interesting cases $0<y<{1\over3}$. Therefore, the previous
equations can be simplified to

$$12\,\kappa_E\,y\,{{\dot a}^2/over{a^2}}\,=\,\mp\,(1\,-\,4\,y\,-\,y^2)\,\rho
\,,\eqno{(5.17a)}$$

$$-\,(1\,-\,4\,y\,-\,y^2)\,a\,{\ddot a}\,+\,y\,(1\,+\,3\,y)\,{\dot a}^2\,=\,
0\,.\eqno{(5.17b)}$$

In terms of the cosmological parameters these equations are rewritten as

$$\Omega\,=\,\mp{4y\over{1-4y-y^2}}\,.\eqno{(5.18a)}$$

$$q\,=\,\pm\,{1\over2}(1\,+3\,y)\,\Omega\,.\eqno{(5.18b)}$$

The only positive root of $(1-4y-y^2)$ is $\sqrt5-2\approx 0.236$. Therefore
we can distinguish two regimes:
I. $0.236<y<{1\over3}$. In this case we must choose the upper sign. The
resulting equations can be applied to the description of the early universe.
II. $0<y<0.236$. In this case we must choose the lower sign. The resulting
equations can be applied to the description of the present universe.

Let us observe that eqs. (5.18a) becomes singular for $y=0.236$. There is no
contradiction here since $\Omega={\rho\over{\rho_c}}$, $\rho_c=3\kappa_E
H^2=3\kappa_E{a^2\over{{\dot a}^2}}$, and for $y=0.236$ we have ${\dot a}=0$
as can be read from eq. (5.17b).

\subsection{5.5}{The Early Universe}

In this case $0.236<y<{1\over3}$ and eqs.(5.18) reduce to

$$\Omega\,=\,-\,{4y\over{1-4y-y^2}}\,.\eqno{(5.19a)}$$

$$q={1\over\,2}\,(1\,+\,3\,y)\,\Omega\,.\eqno{(5.19b)}$$

\nt Let us observe that eq. (5.19b) is the same than that we obtain in
General Relativity, {\it viz.} (C.2a). As in General Relativity one concludes
the existence of a singularity in the past. As explained in the Appendix A,
since matter cannot be compressed beyond the Planck density, it is more
reasonable to consider an initial ball with finite radius. We call this the
"inflationary" stage of our model.

For the early universe matter is described by the state equation
$y={1\over3}$. Then, eqs. (5.19) reduce to

$$9\,\kappa_E\,{{\dot a}^2\over a^2}\,=\,\rho\,,\eqno{(5.20a)}$$

$$a\,{\ddot a}\,+\,3\,{\dot a}^2\,=\,0\,.\eqno{(5.20b)}$$

\nt The solution to eq. (5.20b) is

$$a\,=\,a_0\,{(1\,+\,4\,{{\dot a}_0\over{a_0}})}^{1\slash4}\,\approx\,a_0\,+
\,{\dot a}_0\,t\,.\eqno{(5.21)}$$

\nt In this approximation the horizon radius is

$$r_H(t)\,=\,{a_0\over{{\dot a}_0}}\,(1\,+\,{{\dot a}_0\over{a_0}}\,t)\,ln
(1\,+\,{{\dot a}_0\over{a_0}}\,t)\,.\eqno{(5.22)}$$

Causality is not violated when $r_H(t)>a(t)/c$. This condition is satisfied
for

$$t\,>\,t_{class}\,\approx\,{a_0\over c}\,\approx\,10^{19}\,t_{Planck}\,
\approx\,10^{_24} \,s\,.\eqno{(5.23)}$$

\nt At earlier times quantum mechanical effects dominate the scene. In fact,
the radius of the Universe is exactly the Compton length associated to its
mass. Our classical approach breaks down so that the very concept of
causality is meaningless. Therefore, there is no violation of causality, or
horizon problem.

\subsection{5.6.}{The Present Universe}

In this case $0.236<y<{1\over3}$ and eqs.(5.18) reduce to

$$\Omega\,=\,{4y\over{1-4y-y^2}}\,.\eqno{(5.24a)}$$

$$q\,=\,-\,{1\over2}\,(1\,+\,3\,y)\,\Omega\,.\eqno{(5.24b)}$$

For the present universe matter is described by the state equation $y\approx
0$, i.e., almost pressureless matter. In this case eq. (5.17b) reduces to

$${\ddot a}\,\approx\,0\,.\eqno{(5.25)}$$

Therefore

$$a\,=\,\alpha\,t+\beta\,,\eqno{(5.26)}$$

\nt where $\alpha$ and $\beta$ are integration constants. Therefore, in the
present time the radius of the universe grows linearly with time.
In this case eq. (5.24a) can be inverted to

$$y\,=\,{1\over\Omega}\,[\sqrt{4{(1\,+\,\Omega)}^2\,+\,\Omega^2}\,-\,2\,(1\,+
\,\Omega)]\,.\eqno{(5.27)}$$

Since today matter is almost pressureless small values of $\Omega$ are
favoured by our equation. For the smallest reported value$^7$
$\Omega_{small}=0.01$ we obtain

$$y_{small}\,=\,2.48\,\times\,{10}^{-3}\,.\eqno{(5.28)}$$

\nt This should be compared with the observed value of $p\over\rho$. This can
be determined from the mean random velocity of typical galaxies, $\langle v
\rangle=1\times{10}^3 km/s$, and gives $y_{random}=1\times{10}^{-5}$.
Therefore, our prediction differs by two orders of magnitude with respect to
the observed value. We hope to improve this situation since the estimation of
$y$ from the the random motion of galaxies is a quite rough one. Furthermore,
eq. (5.27) was obtained under the assumption of a time independent state
equation.

\vfill\eject

\subsection{5.7.}{Comments}

The cosmological model we have developed here shows a reasonable agreement
with observational results. In fact, we obtain field equations which among
others: predict an increasing entropy of the universe; are almost consistent
with the observed flatness of the universe; and do not violate causality
(horizon problem). The model is still incomplete in that we have not yet
considered in details the effects of a time dependent state equation for
matter and how this modify the relation (5.18).

The calculation for $p<0$ is almost identical to that for $p>0$. Since one
is used to thinking of the evolution of the universe in terms of $q>0$ this
was the case we favoured in our previous works.$^{10,11,12}$ The entropy is
again an increasing function of time. In this case one has also a linear
growing of the radius of the universe.

\section{6.}{The High-Energy Regime}

Here we explore the high-energy regime of our theory. The form of the
Lagrangian, and of the field equations, for conformal fourth-rank gravity,
puts several strong restrictions, on the kind of matter which can be coupled
consistently to it. The first condition is that matter fields must be
described by conformally invariant Lagrangians in four dimensions. In fact
the trace of eq. (3.27) gives

$$T_4\,=\,G^{\mu\nu\lambda\rho}\,T_{4,\mu\nu\lambda\rho}\,=\,0\,,
\eqno{(6.1)}$$

\nt which is always satisfied by conformal fields.

The situation is similar to that for Einstein gravity in 2 dimensions. The
field equations are

$$\kappa_E\,[R_{\mu\nu}\,-\,{1\over2}\,R\,g_{\mu\nu}]\,=\,T_{\mu\nu}\,.
\eqno{(6.2)}$$

\nt The trace of this equation gives

$$T_2\,=\,g^{\mu\nu}\,T_{2,\mu\nu}\,=\,0\,.\eqno{(6.3)}$$

\nt However, the previous equation collapses to a useless identity since

$$R_{\mu\nu}\,-\,{1\over2}\,R\,g_{\mu\nu}\,\equiv\,0\,.\eqno{(6.4)}$$

\nt In our case, however, this collapse does not occur.

As a second property of our field equations let us observe that the coupling
to conformal fields automatically excludes the existence of a cosmological
constant. In fact, from the Lagrangian

$${\cal L}_4\,=\,\kappa_{CG}\,(\langle R^2\rangle\,+\,\Lambda)\,G^{1\slash4}
\,,\eqno{(6.5)}$$

\nt we obtain

$$\kappa_{CG}\,[ R_{(\mu\nu}\,R_{\lambda\rho)}\,-\,{1\over4}\,(\langle R^2
\rangle\,+\,\Lambda)\,G_{\mu\nu\lambda\rho}]\,=\,T_{\mu\nu\lambda\rho}\,.
\eqno{(6.6)}$$

\nt The trace of this equation is

$$-\,\kappa_{CG}\,\Lambda\,=\,T_4\,,\eqno{(6.7)}$$

\nt but the conformal invariance, eq. (6.1), impose

$$\Lambda\,=\,0\,.\eqno{(6.8)}$$

This situation is again similar to that for Einstein gravity in 2 dimensions.
We have

$${\cal L}_{GR}\,=\,\kappa_E\,(R\,+\,\Lambda)\,g^{1\slash2}\,.\eqno{(6.9)}$$

\nt The field equations are

$$\kappa_E\,[R_{\mu\nu}\,-\,{1\over2}\,(R\,+\,\Lambda)\,g_{\mu\nu}]\,=\,
T_{\mu\nu}\,.\eqno{(6.10)}$$

\nt The trace of this equation gives

$$-\,\kappa_E\,\Lambda\,=\,T_2\,,\eqno{(6.11)}$$

\nt but conformal invariance, eq. (6.3), gives

$$\Lambda\,=\,0\,.\eqno{(6.12)}$$

Therefore the high-energy regime of our theory exhibits all the properties
relevant to a conformal model. In fact it can be consistently coupled to
conformal fields; cf. Ref. 4 for further details. Furthermore, predicts a
cosmological constant which is exactly zero.

\section{7.}{Conclusions}

The results reported here are the product of more than one year effort. We
elaborated many previous versions which were corrected once and again, and
our work was often plagued by false starts.

The conception of new geometries has taught us the importance of observation
in physics and the close relation existing between physics and geometry.

The fourth-rank geometry combined with the observed scale invariance of
physical processes at high-energies leaves us with an almost unique choice
for a gravitational Lagrangian. What we have done here was just to explore
the consequences of this theory.

We would like to emphasize that we did not construct this theory in order to
solve specific problems. We just started from simple principles and explored
their consequences.

In fact, it was unexpected for us to find that the field equations in vacuum,
even when differing from Einstein field equations, give the same solution for
a static spherically symmetric field, namely, the Schwarzschild metric. More
surprising was the fact that our field equations started to differ from those
of General Relativity exactly where they are in disagreement with
observation. It was also unexpected that our field equations provided
solution for long unsolved problems. In fact, entropy is created in the
universe. Causality is not violated, etc.

Of course, from the few results we have presented here one cannot establish
the validity of this theory. It is our purpose to explore further
consequences of our field equations.

Since we began this work with a quotation, it seems convenient to close it in
the same way by including three more quotations.$^{13,14,15}$ Even when they
refer to other historical moments, they can be reread even today with changes
which are obvious. We think they speak by themselves so that no more comments
are necessary.

\bigskip

{\it "The danger of asserting dogmatically that an axiom based on the
experience of a limited region holds universally will now be to some extent
apparent to the reader. It may lead us to entirely overlook, or when
suggested at once reject, a possible explanation of phenomena. The hypothesis
that space is not homaloidal, and again, that its geometrical character may
change with the time, may or may not be destined to play a great part in the
physics of the future; yet, we cannot refuse to consider them as possible
explanations of physical phenomena, because they may be opposed to the
popular dogmatic belief in the universality of certain geometrical axioms- a
belief which has arisen from centuries of indiscriminating worship of the
genius of Euclid."}

\medskip

\rightline{\it W.K. Clifford, 1885}

\bigskip

{\it "[Saccheri's] brilliant failure is one of the most remarkable instances
in the history of mathematical thought of the mental inertia induced by an
education in obedience and orthodoxy, confirmed in mature life by an
excessive reverence for the perishable works of the inmortal dead [Euclid].
With two geometries, each as valid as Euclid's in his hand, Saccheri threw
both away because he was willfully determined to continue in the obstinate
worship of his idol, despite the insistent promptings of his own sane
reason."}

\medskip

\rightline{\it E.T. Bell, 1947}

\bigskip

{\it "People have often tried to figure out ways of getting these new
concepts. Some people work on the idea of the axiomatic formulation of the
present quantum mechanics. I don't think that will help at all. If you
imagine people having worked on the axiomatic formulation of the Bohr orbit
theory, they would never have been lead to Heisenberg's quantum mechanics.
They would never have thought of non-commutative multiplication as one of
their axioms which could be challenged. In the same way, any future
development must involve changing something which people have never
challenged up to the present, and which will we not be shown up by an
axiomatic formulation."}

\medskip

\rightline{\it P.A.M. Dirac, 1973}

\bigskip

\vfill\eject

\entry{Acknowledgements}

This work has been possible thank to the hospitality of the Laboratory of
Theoretical  Physics,  Joint  Institute  for  Nuclear  Research,  Dubna,  the
Istituto di Fisica Matematica "J.-Louis Lagrange",  Universit\`a  di  Torino,
and the International Centre for Theoretical Physics,  Trieste.  One  of  the
authors (A. M.) would like to express his gratitude to Prof.  H.  Boutaleb-J.
for having introduced him to the field of gravitation and cosmology. He would
also like to thank the  Arab  grant  available  for  the  ICTP  Associateship
scheme. The work has been much enriched, at different stages, by  talks  with
M. Ferraris, M. Francaviglia, P. Aichelburg and P. Minning.

\entry{Appendix A. Cosmography}

In this Section we collect the observational results concerning the structure
of the universe and its evolution. Further details can be found in refs. 16
and 17.

The observed isotropy and homogeneity of the universe gives as the only
possible Riemannian geometry for the universe a Friedman-Robertson-Walker
(FRW) geometry. FRW spaces are characterised by the cosmic radius $a(t)$ and
by the constant $k=1,0,-1$, corresponding to a closed, spatially flat, and
open universe, respectively. The curvature properties of a FRW geometry can
be rewritten in terms of the Hubble constant, $H$, and the deacceleration
parameter, $q$. These cosmological parameters can, in principle, be
determined from the observed distance versus velocity Hubble diagram.

At large scales cosmic matter can be described as a perfect fluid which is
characterised by the energy density, $\rho$, and the pressure, $p$, and they
are related by the state equation of matter, ${p\over\rho}=y$. For
$y={1\over3} $ one has a radiation dominated, or ultrarelativistic, perfect
fluid; for $y\approx 0$ one has instead a non-relativistic, or almost
pressureless, perfect fluid.

Associated to the FRW geometry, with the use of the Einstein gravitational
constant, there is a critical density parameter, $\rho_c =3\kappa_E H^2$,
which sets the scale of energy densities. One can then introduce the cosmic
density parameter $\Omega={\rho\over{\rho_c}}$.

The cosmological parameters $H$, $q$ and $\Omega$ are observable, however,
they are quite difficult to determine with accuracy. For the Hubble constant
$H$ there are two preferred values close to 50 and 100 km/sec/Mpc. However
the observed data does not allow to determine the value of the deacceleration
parameter $q$.$^{17}$ According to ref. 17, $\Omega_{obs}\approx 0.1-0.3$,
with an upper safety bound $\Omega_{safe}\approx0.18$. However, early reports
give smaller values such as $\Omega_{small}=0.01$.$^7$

The observed redshift of galaxies shows that the universe is expanding.
Therefore, it was very dense in the past and it is almost diluted today.
Since matter cannot be compressed beyond the Planck density one must consider
the universe as evolving from an initial hot ball. From the conservation of
mass one can furthermore estimate the radius of the initial hot ball to be
$a_0 ={10}^{19}L_{Planck}$.

Further observations, such as the galaxy count-volume test, show that to a
very big extent our universe is spatially quite flat. Therefore, the
parameter $k$ characterising the FRW geometries must be zero, $k_{obs}=0$,
i.e., its geometry is Euclidean.

The last important observation is the fact that the universe is quite
isotropic and homogeneous at large scales. This is an indication that matter
was in causal contact in the very remote past. This condition roughly
translates into $r_H>a$, where $r_H$ is the horizon radius. However, for
very small times, $t<t_{class}\approx{10}^{20}t_{Planck}$, quantum mechanical
effects dominate the scene. In fact, the radius of the universe is exactly
the Compton wavelength associated to its mass and the very concept of
causality is meaningless. Therefore, causality should not be violated only
for $t>t_{class}$.

Any proposed cosmological model must agree with the previously described
observational results. The next task is to develop a gravitational theory
fitting the above observations. The first candidate is General Relativity
(Appendix B).

\entry{A.1. Isotropy, the Cosmological Principle and FRW Spaces}

Observation shows that the universe is isotropic and homogeneous. These
properties give as the only possible Riemannian geometry a FRW metric. In
this case the line element is

$${ds}^2\,=\,{dt}^2\,-\,a^2(t)\,{d\ell}^2\,,\eqno{(A.1)}$$

\nt where

$${d\ell}^2\,=\,{(1\,-\,k\,r^2)}^{-1}\,{dr}^2\,+\,r^2\,{d\Omega}^2\,,
\eqno{(A.2a)}$$

$${d\Omega}^2\,=\,{d\theta}^2\,+\,{sin}^2\theta\,{d\varphi}^2\,.
\eqno{(A.2b)}$$

\nt In the above $a(t)$ is the cosmic scale factor and is interpreted as the
radius of the universe; ${d\ell}^2$ is the line element of a maximally
symmetric three-dimensional space-like section. The radial coordinate r is
written in units such that the constant k takes the values 1, 0 or -1. The
parameter $k$ characterises the geometry of the space-like sections of the
universe. For $k=1$ the universe is closed; for $k=0$ it is flat; for
$k=-1$ it is open.

The Ricci tensor is given by

$$R_{\mu\nu}\,=\,-\,{2\over{a^2}}\,[a\,{\ddot a}\,-\,(k\,+\,{\dot a}^2)]\,
\delta^0_\mu\,\delta^0_\nu\,-\,{1\over{a^2}}\,[a\,{\ddot a}\,+\,2\,(k\,+\,{
\dot a}^2)]\,g_{\mu\nu}\,.\eqno{(A.3)}$$

\nt Hence, the scalar curvature is

$$R\,=\,-\,{6\over{a^2}}\,[a\,{\ddot a}\,+\,(k\,+\,{\dot a}^2)]\,.
\eqno{(A.4)}$$

\nt These quantities can be parametrised in terms of the cosmological
parameters

$$H\,=\,{{\dot a}\over a}\,,\eqno{(A.5a)}$$

\nt which is the Hubble "constant", and it is a true constant only for a de
Sitter space; and

$$q\,=\,-\,{{a{\ddot a}}\over{{\dot a}^2}}\,=\,-\,1-\,{{\dot H}\over{H^2}}\,,
\eqno{(A.5b)}$$

\nt which is the deacceleration parameter.

In a FRW universe the luminosity distance $d_L$ and the redshift $z$ of a
galaxy are related by

$$d_L\,\approx\,{1\over H}\,(z\,+\,{1\over2}\,(q\,-\,1)\,z^2)\,.
\eqno{(A.6)}$$

\nt The distance to a galaxy can be determined by different means. The
redshift is determined by simple spectral techniques. This constitutes the
distance versus redshift Hubble diagram. If $z>0$ one talks of redshifts and
galaxies are receding, while if $z<0$ one talks of blueshifts and galaxies
are approaching. Therefore $H$ can be determined from the slope of the Hubble
diagram while $q$ is related to its convexity.

\entry{A.2. The Matter Content of the Universe. The Perfect Fluid}

In order to be compatible with the observed homogeneity and isotropy of the
universe cosmic matter must be described as a perfect fluid.

A perfect fluid is characterised by the energy-momentum tensor

$$T_{\mu\nu}\,=\,(\rho\,+\,p)\,u_\mu\,u_\nu\,-\,p\,g_{\mu\nu}\,,
\eqno{(A.7)}$$

\nt where $\rho$ and $p$ are the energy density and pressure of cosmic
matter, and

$$u_\mu\,=\,{(g^{00})}^{-1\slash2}\,\delta^0_\mu\,,\eqno{(A.8a)}$$

\nt such that

$$g^{\mu\nu}\,u_\mu\,u_\nu\,=\,1\,.\eqno{(A.8b)}$$

\nt The reduced energy-momentum tensor is

$$S_{\mu\nu}\,=\,T_{\mu\nu}\,-\,{1\over2}\,T\,g_{\mu\nu}\,=\,(\rho\,+\,p)\,
u_\mu\,u_\nu\,-\,{1\over2}\,(\rho\,-\,p)\,g_{\mu\nu}\,.\eqno{(A.9)}$$

In order to relate the energy density $\rho$ and the pressure $p$ one needs a
state equation. Two well understood regimes are the radiation dominated
regime in which $y={p\over\rho}={1\over3}$, and the matter dominated regime
in which $y$ approaches to zero for incoherent matter.

The coupling of gravity to matter needs the Einstein gravitational constant
$\kappa_E$. Combining this constant with the functions characterising the FRW
geometry we obtain the critical density

$$\rho_c\,=\,3\,\kappa_E\,H^2\,=\,1.96\,\times\,{10}^{-29}\,h^2\,g\slash cm^3
\,,\eqno{(A.10)}$$

\nt where

$$h\,=\,{H\over{100\,km\slash sec\slash Mpc}}\,.\eqno{(A.11)}$$

\nt This leads to the introduction of the cosmic energy density parameter

$$\Omega\,=\,{\rho\over{\rho_c}}\,,\eqno{(A.12)}$$

\nt in such a way that $\rho_c$ sets the scale of energy densities.

\entry{A.3. Observed Values of the Cosmological Parameters}

Due to several practical difficulties the observed values of the cosmological
parameters are quite inaccurate. In some cases this inaccuracy does not allow
even to have a reliable value for some parameters. There exists a wide range
of reported values for the cosmological parameters depending on both the
nature of the performed observation and the interpretation of the observed
data. We use the values reported in ref. 17.

\entry{A.3.1. The Hubble Diagram}

In principle, the Hubble diagram should provide, at the same time, the Hubble
constant, $H$, and the deacceleration parameter, $q$. $H$ is related to the
slope of such diagram while $q$ is related to its convexity. However, the
Hubble diagram shows a large dispersion for large values of $H$ such that no
reliable value for $q$ exists today. The reported values are

$$h_{obs}\,=\,0.5\,-\,1.0\,,\eqno{(A.13)}$$

\nt with preferred values closer to 0.5 and to 1.0. The determination of $q$
from deviations from the linear Hubble law is almost imposible with the
present day accuracy of the existing observations.

Since $H$ is positive one can conclude that the universe is expanding. Let
us observe that eq. (A.5a) can be rewritten as

$$a\,=\,{{\dot a}\over H}\,\approx\,{\dot a}\,T\,.\eqno{(A.14)}$$

\nt This equation tells us that the radius of the universe is approximately
its velocity of expansion times the period of expansion. Therefore, $H^{-1}$
can be interpreted as the age of the universe.

\entry{A.3.2. The Energy Density}

The energy density is determined from the cosmic virial theorem and the
infall to the Virgo cluster. According to ref. 17 the observed value for the
energy density ratio is in the range

$$\Omega_{obs}\,\approx\,0.1\,-\,0.3\,.\eqno{(A.15)}$$

\nt There is furthermore a safety upper bound

$$\Omega_{safe}\,\approx\,0.18\,.\eqno{(A.16)}$$

\nt However, early reports$^7$ give smaller values

$$\Omega_{small}\,=\,0.01\,.\eqno{(A.17)}$$

\nt {\bf A.3.3. The Pressure of the Universe}

Up to our knowledge there is no direct determination of the pressure of the
universe. However, an upper bound can be put based on the mean random
velocity, $\langle v\rangle$, of typical galaxies

$${p\over\rho}\,\approx\,{{\langle v\rangle}^2\over{c^2}}\,.\eqno{(A.18)}$$

\nt The proper velocity can be determined from direct measurements and is
given approximately by $\langle v\rangle\approx 1\times{10}^3$ km/s.
Therefore we obtain

$$y_{random}\,\approx\,1\,\times\,{10}^{-5}\,.\eqno{(A.19)}$$

\nt The previous figure put an upper bound to the $p\over\rho$ ratio. It
must be

$${p\over\rho}\,<\,1\,\times\,{10}^{-5}\,.\eqno{(A.20)}$$

\entry{A.3.4. The Radius of the Initial Universe}

The expansion of the universe shows that it has evolved from a very dense
regime in the past and it is almost diluted today. Since matter cannot be
compressed beyond the Planck density it is more reasonable to consider the
universe as evolving from an initial ball.

The radius of the initial universe can be estimated as follows. Let us assume
that the mass of the universe is a conserved quantity. At $t=0$ we assume
that mass was compressed at Planck density. This allows determining $a_0$ to
be

$$a_0\,=\,{[{{M_{Univ}}\over{M_{Planck}}}]}^{1\slash3}\,L_{Planck}\,.
\eqno{(A.21)}$$

\nt The mass of the universe is given by

$$M_{Univ}\,\approx\,{{4\pi}\over3}\,\rho\,{R_{Univ}}^3\,,\eqno{(A.22)}$$

\nt where $R_{Univ}$ is the radius of the universe. This is bounded by the
maximum velocity by which the universe can expand, the velocity of light $c$,
and the time of expansion $H^{-1}$. Therefore

$$R_{Univ}\,\approx\,{c\over H}\,.\eqno{(A.23)}$$

\nt Finally

$$M_{Univ}\,\approx\,{{4\pi}\over3}\,{{\rho c^3}\over H^3}\,\approx\,{10}^
{57}\,M_{Planck}\,.\eqno{(A.24)}$$

\nt Therefore

$$a_0\,\approx\,{10}^{19}\,L_{Planck}\,.\eqno{(A.25)}$$

\nt Our estimation contains an error of a few orders of magnitude which is
not relevant to our analysis.

\entry{A.3.5. Entropy, Flatness and Causality}

{}From the microwave background radiation one observes that the present value
of the total entropy in the universe is so large as to be of order ${10}^{87
}$, in some convenient units.$^{18,19,20}$ On the other hand one would expect
entropy to be governed by the second thermodynamical principle, the statement
that the entropy is always a non-decreasing function of time. One is
therefore faced with the problem of determining whether the entropy of the
universe has always been as large as it is today, $dS=0$, or if it has
evolved from a smaller value. From an intuitive point of view it is quite
improbable that $dS=0$.

Further observations show that our present day universe is almost flat, i.e.,
its geometry is almost Euclidean; this means that

$$k_{obs}\,=\,0\,.\eqno{(A.26)}$$

Another observation concerns the observed isotropy of the universe over large
regions of space; this means that all regions were causally connected in the
past. For this to be the case one should have

$$r_H\,>\,a\,,\eqno{(A.27)}$$

\nt where $r_H$ is the horizon radius which sets the size of the region in
which causal contact can be achieved.

For a FRW space the horizon radius is

$$r_H(t)\,=\,a(t)\,\int^t_0\,{{du}\over{a(u)}}\,,\eqno{(A.28)}$$

\nt which is the maximum distance that light signals can travel during the
age $t$ of the universe.

Let us introduce a time scale

$$t_{class}\,=\,{a_0\over c}\,\approx\,{10}^{19}\,t_{Planck}\,\approx\,
{10}^{-24}\,s\,.\eqno{(A.29)}$$

\nt For times smaller than $t_{class}$ quantum mechanical effects dominate
the scene. In fact, the radius of the universe is exactly the Compton
wavelength associated to its mass. Therefore the very concept of causality is
meaningless. Therefore, causality should be required only for times greater
than $t_{class}$.

\entry{Appendix B. General Relativity}

In General Relativity space-time is conceived as a Riemannian manifold and
the metric $g_{\mu\nu}$ is identified with the gravitational field.

In order to describe the dynamics of the gravitational field we need to
construct an invariant which might be used as Lagrangian. In Riemannian
geometry the simplest invariant which can be constructed is

$$R(g,\,\Gamma)\,=\,g^{\mu\nu}\,R_{\mu\nu}(\Gamma)\,,\eqno{(B.1)}$$

\nt which in the case of a metric space is rewritten as

$$R(g)\,=\,g^{\mu\nu}\,R_{\mu\nu}(g)\,.\eqno{(B.2)}$$

The analytical formulation of General Relativity takes as its starting point
the \break Einstein-Hilbert Lagrangian

$${\cal L}_{EH}(g)\,=\,\kappa_E\,R(g)\,g^{1\slash2}\,,\eqno{(B.3)}$$

\nt where $\kappa_E={c^4\over{8\pi G_N}}$ is the Einstein gravitational
constant; $G_N$ being the Newton constant. The full Lagrangian must consider
also the contributions of matter

$${\cal L}_1\,=\,{\cal L}_{EH}\,+\,{\cal L}_{matter}\,,\eqno{(B.4)}$$

\nt Variation of the Lagrangian with respect to the metric

$${{\delta{\cal L}_1}\over{\delta g^{\mu\nu}}}\,=\,0\,,\eqno{(B.5)}$$

\nt gives the Einstein field equations

$$\kappa_E\,[R_{\mu\nu}(g)\,-\,{1\over2}\,R(g)\,g_{\mu\nu}]\,=\,T_{2,\mu\nu}
\,,\eqno{(B.6)}$$

\nt where

$$T_{2,\mu\nu}\,=\,{1\over g^{1\slash2}}\,{{\delta{\cal L}_{matter}}\over{
\delta g^{\mu\nu}}}\,,\eqno{(B.7)}$$

\nt is the energy-momentum tensor of matter; the 2 stands for the fact that
the energy-momentum tensor is related to Riemannian, second-rank, geometry.

As a starting point for General Relativity one can also consider the
"Palatini" Lagrangian

$${\cal L}_P(g,\,\Gamma)\,=\,\kappa_E\,g^{\mu\nu}\,R_{\mu\nu}(\Gamma)\,g^{1
\slash2}\,.\eqno{(B.8)}$$

\nt In this case one must also consider the contributions of matter

$${\cal L}_2\,=\,{\cal L}_P\,+\,{\cal L}_{matter}\,.\eqno{(B.9)}$$

\nt Now the connection and the metric are varied independently in a procedure
known as the Palatini variational principle. Variation of the Lagrangian with
respect to $\Gamma$ gives

$${{\delta{\cal L}_2}\over{\delta{\Gamma^\lambda}_{\mu\nu}}}\, =\,{{\delta{
\cal L}_P}\over{\delta{\Gamma^\lambda}_{\mu\nu}}}\,+\,{{\delta{\cal L}_{matte
r}}\over{\delta{\Gamma^\lambda}_{\mu\nu}}}\,=\,0\,.\eqno{(B.10)}$$

\nt In all known cases of physical interest one has$^5$

$${{\delta{\cal L}_{matter}}\over{\delta{\Gamma^\lambda}_{\mu\nu}}}\,=0\,.
\eqno{(B.11)}$$

\nt In this case eq. (B.10) reduces to a metricity condition equivalent to
(2.15). Therefore the connection is given by the Christoffel symbol of the
second kind for the metric $g_{\mu\nu}$. Variation with respect to the metric

$${{\delta{\cal L}_2}\over{\delta g^{\mu\nu}}}\,=\,0\,,\eqno{(B.12)}$$

\nt gives

$$\kappa_E\,[R_{\mu\nu}(\Gamma)\,-\,{1\over2}\,g^{\lambda\rho}\,
R_{\lambda\rho}(\Gamma)\,g_{\mu\nu}]\,=\,T_{2,\mu\nu}\,.\eqno{(B.13)}$$

\nt If we now use the previously obtained metricity condition these equations
reduce to the original Einstein field equations (B.6). Therefore, the
procedures of imposing the metricity condition and of applying the
variational principle commute.

Einstein field equations can be rewritten in the Landau form

$$\kappa_E\,R_{\mu\nu}\,=\,S_{2,\mu\nu}\,,\eqno{(B.14)}$$

\nt where

$$S_{2,\mu\nu}\,=\,T_{2,\mu\nu}\,-\,{1\over2}\,T_2\,g_{\mu\nu}\,,
\eqno{(B.15)}$$

\nt with

$$T_2\,=\,g^{\mu\nu}\,T_{2,\mu\nu}\,,\eqno{(B.16)}$$

\nt is the reduced energy-momentum tensor.

Einstein field equations have been applied to many physical situations. The
first classical test of any theory of gravitation is in the solar system. In
this case one needs to solve Einstein field equations in vacuum for a
spherically symmetric field. The solution is the exterior Schwarzschild
metric. Using this metric one can account for the anomalous shift of the
perihelion of inner planets and for the bending of light rays near the solar
surface to an accuracy of 1 per cent or better. In this case one is
describing the effects of the gravitational field alone.

The next test concerns the coupling of gravity to matter. This is achieved,
for instance, when considering the large scale structure of the universe
where gravity becomes coupled to a perfect fluid. As shown in the next
Appendix the agreement with observation is qualitatively good. One obtains
qualitatively good predictions, as the evolution of the universe from an
initial singularity and some good quantitative predictions as the temperature
of the microwave background and the relative abundance of elements. However,
the quantitative agreement is weaker in other aspects. In fact, flatness,
$k_{obs}=0$, implies $\Omega_{GR}=1$, which is hardly observed. Furthermore,
the Standard Model of Cosmology predicts a constant entropy, something which
is difficult to accept on physical grounds. These are some of the reasons to
look for an improved theory for the gravitational field.

Therefore one must consider the possibility that General Relativity is an
incomplete theory. However, it is in good agreement with observation in the
vacuum case: the Schwarzschild solution. Therefore General Relativity
describes well the dynamics of the gravitational field alone, but it fails
when coupled to matter.

Some hints, on how this problem can be approached, come from high-energy
physics. When one tries to quantise General Relativity one discovers that
there are irremovable ultraviolet divergences. This is taken as indicative
that at small distances the geometry of space-time may be different from the
Riemannian one. The current view is that General Relativity, with its
Riemannian structure, is only the low-energy, large distance, manifestation
of a more general theory at small distances. One must therefore construct a
field theory for a more general geometry. The field theory one constructs for
this new geometry must produce, in the absence of matter, a Riemannian
geometry. Furthermore, gravitation, in the form of a theory equivalent to
General Relativity must be recovered. Many possibilities have been explored
mainly in the direction of modifying the affine structure of space-time. Up
to our knowledge, modifications of the metric structure of space-time have
not yet been attempted. As stated in the Introduction, the purpose of this
work is to explore this possibility.

\entry{Appendix C. The Standard Model of Cosmology}

The Standard Model of Cosmology is based on the application of the Einstein
field equations to the universe. They provide the coupling of gravity, or
geometry, given by a FRW metric, to cosmic matter, described by a perfect
fluid.

The Einstein-Friedman equations are equivalent to

$$\rho\,=\,3\,\kappa_E\,{1\over{a^2}}\,(k\,+\,{\dot a}^2)\,>\,0\,,
\eqno{(C.1a)}$$

$$p\,=\,-\,\kappa_E\,{1\over{a^2}}\,[2\,a\,{\ddot a}\,+\,(k\,+\,{\dot a}^2)]
\,.\eqno{(C.1b)}$$

In terms of the cosmological parameters eqs. (C.1) can be rewritten as

$$q\,=\,{1\over2}\,(1\,+\,3\,y)\,\Omega\,,\eqno{(C.2a)}$$

$$\Omega\,=\,1\,+\,{k\over{{\dot a}^2}}\,.\eqno{(C.2b)}$$

The first equation shows that $q>0$ and from here one is used to conceive the
universe as expanding from an initial singularity. This is in agreement with
observation.

However, the Standard Model of Cosmology is in disagreement with some
observations as we will show in detail now.

\entry{C.1. The Entropy Problem}

One question the Standard Model of Cosmology is unable to answer concerns the
problem of the large total entropy in the universe.$^{18,19,20}$ One of the
predictions of the Standard Model of Cosmology is that the expansion of the
universe is an adiabatic isoentropic process. In fact, from the field
equations (C.1) one can easily deduce that

$$T\,dS\,=\,dE\,+\,p\,dV\,=\,d(r\,a^3)\,+\,p\,d(a^3)\,=\,0\,.\eqno{(C.3)}$$

\nt Therefore, the Standard Model of Cosmology predicts that the expansion of
the universe is an adiabatic isoentropic process. There is no entropy
production and the entropy of the universe has always been as large as it is
today, something which is hard to accept based on physical grounds.

\entry{C.2. The Flatness Problem}

The observed flatness of the universe implies $k_{obs}=0$. If we put this
value in eq. (C.2b) we obtain $\Omega_{pred}=1$. However, this is
incompatible with the reported values for $\Omega_{obs}$, (A.14), (A.15) and
(A.16). There exist two possibilities in the face of this impasse. The first
one consists in assuming $k_{obs}=0$, $\Omega_{pred}=1$. This is preferred by
some authors for "aesthetic or philosophical reasons".$^{17}$ This takes us
to the "missing mass" problem. The second possibility is more difficult to
implement. If we accept $\Omega_{obs}<1$, then one should have $k_{pred}=-1$,
as deduced from (C.2a), which corresponds to an open universe. This
possibility is more or less excluded by the cosmological data$^{17}$
indicating that for the large scale structure of the universe $k$ is rather
close to zero. This ambiguous situation is known as the flatness problem.

The above inconsistencies can be removed if the true value of the energy
density parameter $\Omega_{obs}$ is greater than that observed. But dark
matter is hardly observed.

\entry{C.3. The Early Universe}

At early times matter is described by the state equation $y={1\over3}$. In
this case the Einstein field equations reduce to

$$a\,{\ddot a}\,+\,(k\,+\,{\dot a}^2)\,=\,0\,.\eqno{(C.4)}$$

\nt The solution is

$$a\,=\,{(\alpha\,t\,+\,{a_0}^2)}^{1\slash2}\,,\eqno{(C.5a)}$$

$$a\,=\,{(c^2\,t^2\,+\,{a_0}^2)}^{1\slash2}\,,\eqno{(C.5b)}$$

\nt for $k=0,-1$, respectively. The horizon radius are given by

$$r_H\,=\,2\,{c\over\alpha}\,{(\alpha\,t\,+\,{a_0}^2)}^{1\slash2}\,[{(\alpha
\,t\,+\,{a_0}^2)}^{1\slash2}\,-\,a_0]\,,\eqno{(C.6a)}$$

$$r_H\,=\,a_0\,{(1\,+\,x^2)}^{1\slash2}\,ln[x\,+\,{(1\,+\,x^2)}^{1\slash2}]\,
,\eqno{(C.6b)}$$

\nt where $x={{ct}\over{a_0}}$. In both cases causality is not violated for
$t>t_{class}$.

In the Standard Model of Cosmology one usually assumes that $a_0=0$. In this
case causality would be violated. This is called the horizon problem. Our
result differs from the standard one since we have considered a universe
evolving from an initial hot ball with a finite radius rather than from an
initial singularity.

\entry{C.4. Comments}

Hence, it is clear that the observed cosmological data do not fit into the
field equations of the Standard Model of Cosmology, and that even under
strong assumptions on the observed values of the cosmological parameters the
situation cannot be much improved.

We must therefore conclude that the Standard Model of Cosmology is in
disagreement with some cosmological observations. In order to solve the above
problems one can consider inflation.$^{18,19,20}$ Inflation is intended to
solve the entropy, the flatness and the horizon problems. However,
inflationary cosmology will be sound only if later observations will show
that $\Omega_{obs}=1$.

Therefore, our conclusion is that General Relativity is an incomplete theory.
In fact, it describes well, to a very high accuracy, the effects of the
gravitational field alone: the shift of the perihelion of inner planets, the
bending of light in strong gravitational fields, etc., however it fails to
describe the coupling of the gravitational field to matter, for example, the
Standard Model of Cosmology.

We must look therefore for an improved theory for the gravitational field
coinciding with General Relativity in the vacuum case and with a different
way of coupling the gravitational field to matter. The theory of fourth-rank
gravity we have developed satisfies these requirements.

\vfill\eject

\entry{References}

\item{ a.} We thank J. Russo for clarifying us this point.
\item{ b.} The fact of calling $G_{\mu\nu\lambda\rho}$ a metric is a purely
linguistic issue completely unrelated to the mathematical properties of this
object.
\item{ 1.} B. Riemann, {\it \"Uber die hypothesen welche der Geometrie zu
grunde liegen} (1854). This thesis was presented on June 10th, 1854, in
G\"ottingen and it was first published in {\it Abh. K\"onigl. Gesellsch.
Wiss. G\"ottingen} {\bf 13}, 1 (1868). It was translated into English by
W. K. Clifford and published in {\it Nature} {\bf 8}, 14 (1873).
\item{ 2.} H. von Helmholtz, {\it \"Uber die Tatsachen, die der Geometrie zu
Grunde liegen}, {\it Nachr. Ges. Wiss. G\"ottingen} (1868). Reprinted in {\it
Wissenschaftliche Abhandlugen Leipzig} {\bf 2}, 618 (1883).
\item{ 3.} K. F. Gauss, {\it Suppl\'ement a la Th\'eorie de la Combinaison
des Observations}, G\"ottingen (1826).
\item{ 4.} V. Tapia, accompanying paper (1992).
\item{ 5.} C. W. Misner, K. S. Thorne and J. A. Wheeler, {\it Gravitation}
(Freeman, San Francisco, 1973), p. 504.
\item{ 6.} F. Kottler, {\it \"Uber die physikalischen Grundlagen der
Einsteinschen Gravitationstheorie}, {\it Annalen Physik} {\bf 56}, 410
(1918).
\item{ 7.} S. L. Shapiro, {\it Astron. J.} {\bf 76}, 291 (1971).
\item{ 8.} M. Ferraris, private communication (1991).
\item{ 9.} J. L. Synge, {\it Relativity: The General Theory (Interscience
Publishers, New York, 1960).
\item{10.} V. Tapia, preprint IC/92/65, Trieste (1992).
\item{11.} A. L. Marrakchi and V. Tapia, preprint IC/92/86, Trieste (1992).
\item{12.} A. L. Marrakchi and V. Tapia, preprint IC/92/124, Trieste (1992).
\item{13.} W. K. Clifford, {\it The Common Sense of the Exact Sciences}, ed.
by K. Pearson (1885). Reprinted (Dover, New York, 1955).
\item{14.} P. A. M. Dirac, {\it Development of the Physicist's Conception of
Nature}, in {\it The Physicist's Conception of Nature}, ed. by J. Mehra
(Reidel, Dordrecht, 1973).
\item{15.} E. T. Bell, {\it Developments of Mathematics}, (McGraw-Hill, New
York, 1947).
\item{16.} S. Weinberg, {\it Gravitation and Cosmology} (Wiley, New York,
1972).
\item{17.} G. B\"orner, {\it The Early Universe Facts and Fiction} (Springer,
Berlin, 1988).
\item{18.} A. Linde, {\it Rep. Prog. Phys.} {\bf 47}, 925 (1984).
\item{19.} H. Brandenberger, {\it Rev. Mod. Phys.} {\bf 57}, 1 (1985).
\item{20.} A. Linde, {\it Particles Physics and Cosmology} (Harwood Academic
Publishers, Chur, 1990).

\bye